\documentclass[twocolumn]{aa} 
\usepackage{graphicx}
\usepackage{natbib}

\usepackage{txfonts}
\begin{document}
    \title{High Resolution X-ray Spectroscopy of T Tauri Stars in the Taurus-Auriga
   Complex}


   \author{Alessandra Telleschi\inst{1}, Manuel G\"udel\inst{1}, Kevin~R. Briggs\inst{1}, Marc Audard\inst{2}
          \and
          Luigi Scelsi\inst{3}}
   \authorrunning{A. Telleschi et al.}
   \titlerunning{High Resolution X-ray Spectroscopy of TTS in Taurus-Auriga}
   \offprints{A. Telleschi}

   \institute{Paul Scherrer Institut, W\"urenlingen and Villigen,
              CH-5232 Villigen PSI, Switzerland\\
              \email{atellesc@astro.phys.ethz.ch}
              \and
             Columbia Astrophysics Laboratory, Mail code 5247,
             550 West 120$^{th}$ Street, New York, NY 10027
              \fnmsep\thanks{\emph{New address (since September 2006): }Integral Science Data Centre, Ch. d'Ecogia 16, CH-1290 Versoix, Switzerland \& Geneva Observatory, University of Geneva, Ch. des Maillettes 5
1, 1290 Sauverny, Switzerland}
\and
             Dipartimento di Scienze Fisiche ed Astronomiche, 
             Sezione di Astronomia, Universit\`a  di Palermo, 
             Piazza del Parlamento 1, 90134 Palermo, Italy
             }

   \date{Recived 2006; accepted 2006}

 \abstract
   {Differences have been reported between the X-ray emission of 
   accreting and non-accreting stars. Some observations have suggested that accretion 
   shocks could be responsible for part of the X-ray emission
   in Classical T Tauri stars (CTTS).}
    { We present high-resolution X-ray spectroscopy of nine pre-main sequence stars
    in order to test the proposed spectroscopic differences between accreting and
    non-accreting pre-main sequence stars.}
    {We use X-ray spectroscopy from the {{\it XMM-Newton}} Reflection Grating Spectrometers and 
    the EPIC instruments. We interpret the spectra using optically thin thermal models with 
    variable abundances, together with an absorption column density. For BP Tau
    and AB Aur we derive electron densities from the  O\,{\sc vii} triplets.}
   {Using the O\,{\sc vii}/O\,{\sc viii} count ratios as a diagnostic 
    for cool plasma, we find that CTTS display a soft excess
    (with equivalent electron temperatures of $\approx 2.5-3$ MK) when compared with WTTS or 
    zero-age main-sequence stars. Although the O\,{\sc vii} triplet in BP Tau is consistent  
    with a high electron density ($3.4 \times 10^{11}$~cm$^{-3}$), we find a low
    density for the accreting Herbig star AB Aur ($n_{\rm e}< 10^{10}$~cm$^{-3}$). 
    The element abundances of accreting and non-accreting stars are  similar. The Ne 
    abundance is found to be high (4-6 times the Fe abundance)
    in all  K and M-type stars. In contrast, for the three G-type stars 
    (SU Aur, HD 283572, and HP Tau/G2), we find an enhanced Fe
    abundance (0.4-0.8 times solar photospheric values) compared to later-type stars. }
   {Adding the results from our sample  to former high-resolution
    studies of T Tauri stars, we find a soft excess in all accreting stars,
    but in none of the non-accretors.
    On the other hand, high electron density and high Ne/Fe abundance ratios 
    do not seem to be present in  all accreting pre-main sequence stars.}


   \maketitle
%
\section{Introduction}

Young, late-type stars are known to be luminous X-ray emitters. This is particularly
true for T Tauri stars, which are divided  into two classes: the Classical T Tauri stars
(CTTS) and the Weak-line T Tauri stars (WTTS). The CTTS display strong H$\alpha$
emission lines as a signature of accretion and an infrared excess as a signature
of the presence of a circumstellar disk.
The WTTS display much weaker H$\alpha$ lines
and no infrared excess, a sign that accretion has ceased and a thick disk is
no longer present. While CTTS are thought to be in an earlier evolutionary stage,
both types of stars are simultaneously present in star forming regions.

X-rays from T Tauri stars were first detected in early studies
with the {\it Einstein} satellite \citep{feigelson81} and surveyed
more extensively with the {\it ROSAT} satellite (e.g., \citealt{feigelson93, 
neuhaeuser95}).  Both types of young stars exhibit variability on time 
scales of order of hours to days. Their overall X-ray 
properties were found to be similar to X-ray characteristics
of main-sequence stars, and the X-ray emission was interpreted
to arise from a corona.

However, most young stellar objects, especially in their early evolutionary stage,
are believed to be fully convective, therefore the generation of
magnetic fields through a solar-like dynamo should not be possible.
Also, the presence of a disk and accretion in CTTS adds another important
element: shocks generated by accretion of circumstellar
material onto the star could contribute to the X-ray emission.

The influence of accretion on X-ray emission processes has been debated.
In the {\it ROSAT} observations of the Taurus-Auriga complex, \citet{stelzer01}
found that the $L_{\rm X}/ L_{*}$ ratio (where  $L_{*}$ is the bolometric luminosity of the star)
is lower in CTTS than in WTTS.
This has been confirmed in the recent, more comprehensive
{\it XMM-Newton Extended Survey of the Taurus Molecular Cloud} (XEST),
but on the other hand, no direct correlation is found between accretion
rate and $L_{\rm X}/L_{*}$ \citep{telleschi06b}. Similar results have also been 
reported from the {\it Chandra} Orion Ultradeep Project (COUP), where the CTTS 
are found to be less luminous in X-rays than WTTS \citep{preibisch05}. In both 
surveys, the bulk X-ray emission is consistent with coronal
emission.

On the other hand, recent results from high-resolution X-ray spectroscopy 
of some CTTS seem to indicate the presence of X-ray emission generated in
accretion shocks. The most evident case is TW Hya. The {\it Chandra} and 
{\it XMM-Newton} spectra of TW Hya \citep{kastner02, stelzer04} are 
dominated by emission from plasma at temperatures of $\approx 3$~MK 
which is much lower than usually found in T Tauri stars.
The O\,{\sc vii} triplet, consisting of lines at 21.6 \AA~(resonance, $r$ line), 
21.8 \AA~ (intercombination, $i$ line), and 22.1 \AA~(forbidden, $f$ line), displays
an $f/i$ flux ratio much below unity. This triplet is density-sensitive \citep{gabriel69}; in late 
type stars, typical values for the $f/i$ ratio are larger than unity  \citep{ness04,testa04}, 
indicating densities of at most a few times 10$^{10}$ cm$^{-3}$. In TW Hya, the measured 
electron density exceeds $10^{12}$ cm$^{-3}$ \citep{kastner02, stelzer04}, 
i.e., it is at least two orders of magnitude larger than typical coronal densities. 
Studies of other density-sensitive lines lead to similar results: the He-like Ne\,{\sc ix} 
triplet in the {\it Chandra} observations gives $\log n_e = 12.75$ \citep{kastner02}, 
while a study by \citet{ness05} presents additional evidence for high density from 
the flux ratios of  Fe\,{\sc xvii} lines.
Further, \citet{kastner02} and \citet{stelzer04} found an abundance anomaly in the spectrum of TW Hya,
with certain metals being underabundant with respect to the solar photospheric abundances 
while nitrogen and neon are found to be strongly overabundant. This anomaly was 
interpreted  by \citet{stelzer04} as being due either to metal depletion by condensation onto grains 
in the accretion disk or to an abundance anomaly present in the original molecular cloud.
A particularly high Ne/Fe abundance ratio was also found by \citet{argiroffi05} 
for TWA 5, a stellar system in the TW Hya association. This object is a quadruple system 
and one of the components could be an accreting star \citep{mohanty03}, although it is not 
possible to identify which of the stars is the X-ray source. 
The high Ne/Fe ratio could be due to the accreting star, 
but could also be an environmental effect of the TW Hydrae association 
\citep{argiroffi05}. \citet{drake05b} have proposed to use the Ne/O abundance
ratio as a diagnostic for metal depletion in accreting 
stars. This ratio is in fact found to be substantially larger in TW Hya 
than in all other studied stars (also higher than in TWA 5, another
star of the TW Hydrae association).   

\citet{schmitt05} have discussed the high resolution X-ray spectrum of the
CTTS BP Tau observed with {\it XMM-Newton}.
This spectrum also displays an $f/i$ ratio smaller than unity, resulting
in an electron density of $\log n_e \approx 11.5$. However, the BP Tau spectrum is
dominated by a hot plasma component. The authors interpreted the hotter component
as originating from a corona or from magnetic interaction between disk and corona,
while the soft component may originate from accretion shocks.

\citet{robrade06} have presented a comparative study of high resolution spectra
of four pre-main sequence accreting stars: BP Tau, CR Cha, SU Aur, and TW Hya.
They tentatively added CR Cha as another CTTS with a low O\,{\sc vii} $f/i$ ratio,
 but the low signal-to-noise (S/N) ratio makes the measurements uncertain.
The O\,{\sc vii} triplet was not detected in SU Aur.
The Ne abundance is found to be enhanced relative to Fe also in CR Cha and
BP Tau. However, the Ne/O abundance ratio of BP Tau is similar to the ratio found for
other (non-accreting) stars and not as high as that of TW Hya,
which may be due to evolutionary effects in the accretion disk \citep{drake05b}.

Recently, \citet{guenther06} studied the {\it Chandra} spectrum of 
another CTTS, V4046 Sgr, and measured a high O\,{\sc vii} $f/i$ ratio,
consistent with electron densities of $\log n_e \approx 11.5$.

High resolution spectroscopy is of crucial importance to identify the soft component
in X-ray spectra of young pre-main-sequence stars because emission lines formed at low coronal
temperatures can be accessed, such as lines of C\,{\sc v}, C\,{\sc vi}, N\,{\sc vi}, 
N\,{\sc vii}, O\,{\sc vii}, and O\,{\sc viii}, whereas the energy resolution of CCDs 
is insufficient to reveal these lines individually. 
The possibility to measure density-sensitive lines is also essential to distinguish 
between X-ray emission that may originate in accretion shocks and coronal emission.
A comparative study of X-ray spectra of CTTS and WTTS should further our understanding
of the role of accretion in the production of X-ray emission.

In this paper we present high resolution X-ray spectra of four CTTS, four WTTS,
and a one Herbig Ae/Be star.
These observations are part of the XEST project \citep{guedel06b}, 
the {\it XMM-Newton} survey of the most populated regions in the Taurus Molecular Cloud.
The purpose of that survey is to study X-ray emission in a large fraction
of the TMC population.
While the spectra of  HD 283572, SU Aur, and BP Tau have been extracted from archival data and have
already been presented in the literature \citep{scelsi05, schmitt05, robrade06},
the RGS spectra of the other stars are shown for the first time in this paper.
We reanalyze the spectra of the former three stars to provide a consistent 
comparison.

The structure of our paper is as follows. The stellar sample is described in
Section~\ref{sample}, while we present our observations
and data reduction in Sect.~\ref{observations}. The results are presented in
Sect.~\ref{results}, while a detailed discussion of the O\,{\sc vii} triplets is
presented in Sect.~\ref{triplet}. In Sect.~\ref{discussion} we discuss our results
and  Sect.~\ref{conclusions} contains our conclusions.


\section{Stellar Sample}\label{sample}

We present RGS spectra of 9 pre-main sequence stars. Four of them are
CTTS, four are WTTS, while one is an intermediate-mass young star, belonging
to the class of Herbig stars \citep{herbig60}. The principal properties
of the stars are listed in Table~\ref{prop}. We discuss some specific characteristics below.

\begin{table*}
\caption{Properties of target stars. $L_*$ is the stellar (photospheric) bolometric luminosity, $M$ is the mass, $A_{\rm V}$ and $A_{\rm J}$
         are the extinctions in the $V$ and $J$ bands, respectively, $P_{\rm rot}$ is the rotation period, EW(H$\alpha$) is
         the equivalent width of the H$\alpha$ line  (positive for emission), and $\dot{M}$ is the mass accretion rate. If not otherwise
         noted, the properties are reported from \citet{guedel06b} and references therein.}             
\label{prop}      
\centering                          
\begin{tabular}{l c c c c c c c c c c c c}        
\hline\hline                 
Star & XEST No.$^{a}$ & $L_*$ & $M$ & Age & $R$ & $A_{\rm V}$ & $A_{\rm J}$ & $P_{\rm rot}$ & Spec. & EW(H$\alpha$)& $\log \dot{M}$ & Type\\    
 &  & ($L_{\odot}$) & ($M_{\odot}$) & (Myr) & ($R_{\odot}$) & (mag) & (mag) & (d) & &(\AA) & ($M_{\odot}$~yr$^{-1}$) & TTS  \\    
\hline                        
 HD 283572 & 21-039 & 6.50 & 1.70 & 7.92 & 2.56 & 0.38 & 0.11 & 1.55 & G5 & 0 & -- & W\\
 V773 Tau  & 20-042 & 5.60 & 1.53 & 6.35 & 1.91 & 1.39 & 0.31 & 3.43 & K2 & 4-10 & $< -10$ & W\\
 V410 Tau$^c$ &  23-032/24-028  & 2.20 & 1.51 & 2.74 & 2.31 &0.67 & 0.00 & 1.94 & K4 & 2-3 & $< -8.8$ & W\\
 HP Tau/G2 & 08-051 & 6.50 & 1.58 & 10.5 & 2.34 &  2.08 & 0.66 & 1.20 & G0 & 0-5 & -- & W\\
 SU Aur    & 26-067 & 9.90 & 1.91 & 6.02 & 3.06 & 0.90 & 0.21 & 1.70 & G2 & 2-6 & -8.30/-8.20 & C\\
 DH Tau    & 15-040 & 0.56 & 0.47 & 1.53 & 1.82 & 1.25 & 0.32 & 7.00 & M1 & 39-72 & -8.95/-8.30 & C\\
 BP Tau    & 28-100 & 0.95 & 0.75 & 1.91 & 1.97 & 0.49 & 0.14 & 7.60 & K7 & 40-92 & -7.88/-7.54 & C \\
 DN Tau    & 12-040 & 1.00 & 0.56 & 1.05 & 2.25 & 0.49 & 0.34 & 6.30 & M0 & 12-87 & -8.73/-7.79 & C\\
 AB Aur    & 26-043 & 49.0& 2.70 & 4.0 & 2.31 & 0.25 & 0.24 & $<$1.46& B9.5-A0 & 22-44 & $\approx -8^{b}$& Ae\\
\hline                                   
\end{tabular}
\begin{minipage}{0.97\textwidth}
\footnotetext{
\hskip -0.5truecm $^a$ XEST catalog number: the first two digits define the exposure numbers, the last three digits are the source numbers \citep{guedel06b}.\\
$^b$ From \citet{telleschi06a} and references therein.\\
$^c$ V410 Tau was observed in two different exposures, therefore two different XEST numbers are given (see below).}
\end{minipage} 
\end{table*}

HD 283572 is a single G type star that does not show signs of accretion
and is therefore classified as a WTTS  \citep{kenyon98}.
Also, there is no evidence for an infrared excess that would
be due to a circumstellar dust disk \citep{furlan06}. 
\citet{favata98} have studied its X-ray variability,
analyzing data from {\it Einstein}, {\it ASCA}, {\it ROSAT}, and SAX. 
Results on the {\it XMM-Newton} data set have already been presented
by \citet{scelsi05}.

V773 Tau is a system composed of four stars.
The system was detected to be a binary by \citet{ghez93} and \citet{leinert93};
the primary component was then discovered to be a binary itself
\citep{welty95}.
Recently, a fourth component has been detected
by \citet{duchene03}.
These authors showed that the two close components A and B 
reveal properties of WTTS, while the C component
shows a near infrared excess typical for CTTS.
The fourth component has been classified as an infrared companion (IRC).
The nature of IRC is debated: they could be
deeply embedded TTS undergoing strong accretion (see e.g. \citealt{koresko97})
or they may be embedded protostars (e.g. \citealt{ressler01}).
In the latter case, the four stars would be in three different evolutionary
phases. 
We refer the reader to \citet{duchene03} for an exhaustive
discussion.
In this work we will treat V773 Tau as a WTTS for two reasons:
the primary and secondary stars (the most luminous components) are WTTS, and the
hydrogen column density ($N_{\rm H}$) found in our spectral fits (2.0 $\times 
10^{21}$~cm$^{-2}$) is consistent with the optical extinctions found for 
the two WTTS ($A_{\rm V}$ = 1.39 mag) if a standard gas-to-dust ratio is assumed 
\citep[$N_{\rm H}/A_{\rm V} = 2 \times 10^{21}~{\rm cm}^{-2}~{\rm mag}^{-1}$;][]{vuong03}. 
The masses of V773 Tau A and B are 1.5 and $\approx 1 M_{\odot}$, respectively,
while the mass of the C component is only $\approx 0.7  M_{\odot}$. The IRC
would have a mass of $\le 0.7 M_{\odot}$.
Because the X-ray luminosities of TTS are correlated with mass \citep{preibisch05,telleschi06b},
we expect the A+B components to dominate the X-ray spectrum.
X-rays from V773 Tau were previously detected with {\it ROSAT} \citep{feigelson94} and with
{\it ASCA} \citep{skinner97}. 

V410 Tau is a triple system \citep{ghez93, ghez97}. 
V410 Tau A and B are separated by $0.12 \arcsec$ with B being much fainter.
The C component is also very much fainter than V410 Tau A.
The primary is of spectral type  K4 with a mass of 1.5 $M_{\odot}$, 
and has an age of 2.74 Myr. 
V410 Tau was detected in X-rays by {\it ROSAT} \citep{strom94}, and
more recently \citet{stelzer03} presented a set of {\it Chandra} 
observations of this source.

SU Aur is a single star of spectral type G2, with a mass of 1.9 $M_{\odot}$.
The properties of this object are described by \citet{dewarf03}.
Despite the low equivalent width (EW) of the H$\alpha$ line
reported in Table~\ref{prop}, its early spectral type and the measured infrared excess classifies SU Aur
as a CTTS \citep{muzerolle03}. The source is one of the X-ray brightest CTTS and
was already detected by \citet{feigelson81} with the
{\it Einstein Observatory}, and by \citet{skinner98}
in an {\it ASCA} observation.
Results from its {\it XMM-Newton} spectrum
have recently been published by \citet[][ further analysis will be presented 
by Franciosini et al. 2007, in preparation]{robrade06}. 

HP Tau/G2 is a G0 star that forms a triple system with HP Tau/G3 
(separation 10$\arcsec$), the latter itself being a binary \citep{richichi94}. 
The X-ray spectrum of HP Tau/G2 is contaminated by HP Tau/G3. However,
fitting separate PSF in the EPIC data, we found that  
the source counts of HP Tau/G3 amount to only about 8\% of the
count rates of HP Tau/G2 \citep{guedel06b}. The contamination due to HP Tau/G3 is 
therefore negligible. 
Bipolar outflows have been detected from HP Tau/G2 \citep{duvert00},
although the star is classified as a WTTS.

DH Tau is another binary, the primary DH Tau A being a CTTS, with a mass of 0.5
$M_{\odot}$  and an age of 1.5 Myr. DH Tau B is a brown dwarf
\citep{itoh05} with mass of 0.03-0.04 $M_{\odot}$ and an age of 3-10 Myr.

BP Tau is a CTTS of spectral type K7 with a mass of 0.75 $M_{\odot}$ and an
age of $\approx$ 2 Myr, according to the evolutionary model of \citet{siess00}.
The star was already observed by the {\it Einstein Observatory}, with 
$L_{\rm X} \approx 1 \times 10^{30}$~erg s$^{-1}$ \citep{walter81}.
Results from the {\it XMM-Newton} spectrum have been discussed
by \citet{schmitt05} and \citet{robrade06}. They measured a relatively high
density in the O\,{\sc vii} triplet, suggesting
that some of the X-rays could be produced in accretion shocks.

DN Tau is a CTTS of spectral type M0 with a mass of
0.56 $M_{\odot}$ and an age of $\approx$ 1 Myr. According to
\citet{muzerolle03}, the inclination angle is 35$^\circ$, and the disk
is relatively small (0.05 AU in radius).

AB Aur is a Herbig star of spectral type B9.5-A0 with a mass of
2.7 $M_{\odot}$ and an age of $\approx 4$ Myr
(DeWarf \& Fitzpatrick, private communication). It is
surrounded by a disk from which it is accreting material (see for 
example \citealt{catala99}): it is therefore
useful to compare the X-ray emission of this star with the
X-ray emission of CTTS. A detailed discussion  of the X-ray
spectrum of this star is presented in \citet{telleschi06a}.

\begin{table*}
\caption{Observation log}             
\label{log}      
\centering                          
\begin{tabular}{l c c c c c c c }        
\hline\hline                 
Stars & XEST No. & ObsID$^{1}$ & Instruments & EPIC  & Start Time & Stop Time & Exposure  \\    
 &  &  &  & Filter  & y-m-d h:m:s & y-m-d h:m:s & [s]\\    
\hline                        
 HD 283572 & 21-039 & 0101440701 & RGS2, MOS1  &  Medium & 2000-09-05 02:57:44 & 2000-09-05 15:47:55 & 46211\\
 V773 Tau & 20-042 & 0203542001 & RGS1, RGS2, MOS1 &  Medium & 2004-09-12 07:04:43 & 2004-09-12 15:52:37& 31674\\
 V410 Tau & 23-032& 0086360301 & RGS1, RGS2, MOS1 &  Medium & 2001-03-11 12:46:45 & 2001-03-12 09:13:24& 73599\\
 V410 Tau & 24-028& 0086360401 & RGS1, RGS2, MOS1 &  Medium & 2001-03-12 09:29:38 & 2001-03-12 21:47:51& 44293\\
 HP Tau/G2 & 08-051 & 0203540801 & RGS1, RGS2, MOS1 &  Medium & 2004-08-26 06:36:23 & 2004-08-26 18:10:59& 41676\\
 SU Aur    & 26-067 & 0101440801 & RGS1, RGS2, MOS1 &  Thick & 2001-09-21 01:34:17 & 2001-09-22 13:34:31& 129614\\
 DH Tau    & 15-040 & 0203541501 & RGS1, RGS2, MOS1 &  Medium & 2005-02-09 13:12:40 & 2005-02-09 22:38:18& 33938\\
 BP Tau    & 28-100 & 0200370101 & RGS1, RGS2, MOS1 &  Thick & 2004-08-15 06:14:30 & 2004-08-16 18:42:57 & 131307\\
 DN Tau    & 12-040 & 0203542101 & RGS1, RGS2, MOS1 &  Medium & 2005-03-04 20:22:29 & 2005-03-05 05:01:54& 31165 \\
 AB Aur    & 26 043 & 0101440801 & RGS1, RGS2, MOS2 &  Thick & 2001-09-21 01:34:17 & 2001-09-22 13:34:31& 129614\\
\hline                                   
\multicolumn{5}{l}{$^1$ {\it XMM-Newton} observation identification number}\\
\end{tabular}
\end{table*}


\section{Observations and data analysis}\label{observations}

The XEST campaign collected 28 {\it XMM-Newton} observations 
within the Taurus-Auriga complex \citep{guedel06b}. 
We obtained high-resolution X-ray spectra from nine targets within these
fields of view, which are in the focus of this paper.
The observation log is given in Table~\ref{log}.
For each star, we used each spectrum from the two Reflecting Grating Spectrometers (RGS,
\citealt{denherder01}) and one of the spectra from the MOS-type  European Photon
Imaging Cameras (EPIC, \citealt{turner01}).
For HD 283572, the RGS1 instrument was  out
of operation and, therefore, we restrict our analysis to the RGS2 and
MOS1 instruments. For all observations, the MOS
instrument observed in full frame mode.
Its detectors are sensitive in the energy range of 0.15--15.0 keV with
a spectral resolving power of $E/\Delta E = 20-50$. The RGSs are suited for
high-resolution spectroscopy, operating in the wavelength range of 6--35~\AA~
with a resolution of ~$\Delta \lambda \approx 60-76~$m\AA.

We reduced our data using the Science Analysis System (SAS) version 6.1.
For the EPIC detectors, the data were reduced using the {\it emchain}
task and the sources were detected using the maximum likelihood algorithm
{\it emldetect} (see \citealt{guedel06b} for more details). We reduced the RGS
data using the  task {\it rgsproc}.
The calibration files used in the data reduction are those described
by \citet{pollock04}.
We extracted the total (source+background) 
spectra and the background spectra separately. In order to optimize
the signal-to-noise (S/N) ratio of the spectra and given the weakness of
most of them, we decided to include only 85\% of the cross-dispersion Point
Spread Function (PSF, {\it xpsfincl=85}). For the background extraction, we kept the
exclusion region of the cross-dispersion PSF and the inclusion region
of the pulse-height distribution at the default values, namely
95\% and 90\%, respectively ({\it xpsfexcl=95} and
{\it pdistincl=90}).

We fitted the spectra in XSPEC \citep{arnaud96}, using the optically-thin
collisionally-ionized plasma model calculated with the Astrophysical Plasma
Emission Code (APEC, \citealt{smith01}). In order to account for calibration
discrepancies between the RGS and the MOS detectors, we introduced effective 
area factors fixed at 1.0 for the MOS and 1.05 for the RGS, applicable to
our wavelength region according to \citet{kirsch04}.

Because we are mainly interested in the line-dominated RGS spectra, we used, for the fit
procedure, both RGS spectra (if available) but only one MOS
spectrum (MOS2 for AB Aur and MOS1 for all other stars), the latter confined to short wavelengths (between 1.5 and 9.35 \AA).
This range is important to obtain abundances of Mg, Si, S, and Fe. For the RGS spectrum,
we used the wavelength region between 8.3 \AA~and 25.0 \AA, except for AB Aur, where
we fitted RGS1 between 10 \AA~ and 28 \AA~ and RGS2 between 8.3 \AA~and 26.5 \AA, in order to include the
N\,{\sc VII} line at 24.78 \AA~(we choose slightly different wavelength intervals in 
order to exclude the spectral ranges where an accurate background subtraction was most difficult).
A similar approach has  been followed by \citet{audard03} and \citet{telleschi05}.

In order to use the $\chi^2$ statistic for the fitting procedure,
the total spectra were binned to a minimum of 20 counts per bin for RGS
(15 counts per bin for the faint spectrum of DN Tau) and to a minimum of 15 counts per
bin for MOS.

The spectra were analyzed using two different approaches.
First, we fitted the data with a model consisting of a continuous
emission measure distribution (EMD) approximated by two power laws
as used in the fits to the EPIC spectra in the XEST survey
\citep{guedel06b}.
The EMD model is approximated by a grid of isothermal
components, defining two power laws, one at low temperatures
and one at high temperatures. This model was motivated
by the EMD shape found from previous high resolution X-ray spectroscopy of
young solar analog stars \citep{telleschi05}. The model can be described
by
 \begin{equation}\label{eq:dem}
Q(T) = \left\{ \begin{array}{ll} EM_0 \cdot (T / T_0)^{\alpha}\quad   &, {\rm for} ~ T \le  T_0 \\
                                      EM_0 \cdot (T / T_0)^{\beta}\quad     &, {\rm for} ~ T >  T_0 \end{array} \right.
\end{equation}
where $T_0$ is the temperature at which the power laws cross,
and $EM_0$ is the emission measure per $\log T$ at this temperature.
The slopes of the power laws below and above $T_0$ are $\alpha$ and
$\beta$, respectively. We left $\beta$ free to vary (between $-3 \le \beta \le 1$),
while we fixed the slope $\alpha$ at a value of 2, as suggested from
EMDs derived in previous studies \citep{telleschi05, argiroffi04}.
We set low and high temperature cut-offs of the two power laws
at $\log T = 6.0$ and $\log T = 8.0$. The parameters that
we fitted hence are $T_0$, $EM_0$, $\beta$ and a selection of elemental
abundances.

\begin{table*}
\caption{Target stars: Results from the EPIC spectral fits.
$L_{\rm X}$ is calculated in the range 0.3-10 keV}             
\label{epicfit}      
\centering                          
\begin{tabular}{cc c c c c c}        
\hline\hline                 
Stars & $N_{\rm H}$ & $T_0$ & $\beta$ &  EM$^1$ &  T$_{\rm av}$ & $L_{\rm X}$  \\    
   & ($10^{22}~\rm{cm}^{-2}$) & (MK) &   & ($10^{52}~\rm{cm}^{-3}$) & (MK) &  ($10^{30} \rm{erg~s^{-1}}$)  \\    
\hline                        
 HD 283572 & 0.08 (0.07, 0.08)& 10.4 (10.0, 10.7)& -0.94 (-1.02, -0.87)&  114.3 & 14.4 & 13.0 \\
 V 773 Tau & 0.17 (0.17, 0.17)& 8.7  (8.4, 9.1)& -0.87 (-0.93, -0.82) & 89.8 & 13.0  & 9.5  \\
 V 410 Tau & 0.02 (0.02, 0.03)&10.4 (9.4, 10.8) & -1.17 (-1.28, -0.96)& 44.5 & 13.2 & 4.7 \\
 HP Tau/G2 & 0.41 (0.39, 0.42) & 9.2 (8.6, 9.9) & -1.37 (-1.46, -1.28) & 94.6 & 11.0 & 9.7  \\
 SU Aur    & 0.47 (0.43 0.48) & 6.4 (6.2, 7.6) & -1.11 (-1.21, -1.06)& 95.4 & 8.9 & 9.5  \\
 DH Tau    & 0.20 (0.19, 0.21)& 11.5 (11.0, 12.1)& -1.38 (-1.49, -1.26) & 80.6 & 13.4 & 8.5  \\
 BP Tau    & 0.06 (0.06, 0.07) & 7.1 (6.7, 7.6) & -0.67 (-0.73, -0.60)& 12.8 & 12.3 &1.4  \\
 DN Tau    & 0.07 (0.07, 0.08) & 10.5 (9.5, 11.5)& -1.25 (-1.51, -1.02)& 11.0& 12.9 &1.1  \\
 AB Aur    & 0.05 (0.03, 0.07) & 4.8 (4.3, 5.5)& -1.55 (-1.78, -1.41)& 3.8   & 5.7 & 0.4 \\
\hline                                   
\multicolumn{7}{l}{ $^1$ Total EM integrated over temperature bins between $\log T = 6 -7.9$ [K] (see \citealt{guedel06b} for
more details)}\\
\end{tabular}
\end{table*}

Although our EMD model is constrained by some approximations (double power-law
shape, $\alpha$ fixed at +2), for a coronal source it is physically more plausible than a 
2 or 3-temperature model, as the emission is described by a continuous 
set of temperatures rather than by only 2 or 3 isothermal components. However,
because of the approximations, the EMD model could be inadequate if the 
X-rays were produced, for example, in near-isothermal
accretion shocks. Therefore, we further test our results with a model
with two or three isothermal components.
We also apply  an EMD model in which $\alpha$ is a free parameter.

In both models, we generally fixed the absorption column density $N_{\rm H}$ at the value
found in the XEST survey from the EPIC data (Table~\ref{epicfit}). These values agree well with
$A_{\rm V}$ measurements from the literature (see Table~\ref{prop}) if a
standard gas-to-dust ratio is assumed. 
 We decided to fit $N_{\rm H}$ only for the high-quality spectrum of SU Aur. 
The value found in the XEST survey in fact led to problems with some RGS 
lines, which was not the case when we fitted $N_{\rm H}$.
The $N_{\rm H}$ found in the fits of SU Aur agrees much better with the
$A_{\rm V}$ values given in the literature ($A_{\rm V} = 0.9$, \citealt{kenyon95}).

We fitted elemental abundances for the lines that are clearly seen in the
spectra (O, Ne, Fe, Mg, Si, and S; also N for AB Aur), while we fixed the abundances of those
elements that do not show significant features (C, N, Ar, Ca, Al, and Ni; S for AB Aur)
at the values used in the XEST data analysis (C=0.45, N=0.788, Ar=0.55, Ca=0.195, Al=0.5,
Ni=0.195, and S=0.417; see \citealt{guedel06b}). All abundances are calculated 
with respect to the solar photospheric abundances of \citet{anders89} except
Fe, for which we used the value given by  \citet{grevesse99}. The above coronal abundances are 
characteristic for active, young stars  described  in the literature 
\citep{garciaalvarez05,argiroffi04}. They are arranged according to a weak 
``inverse First Ionization Potential (FIP) effect'' (elements with
higher FIP are overabundant relative to low-FIP elements if normalized to solar
photospheric abundances; \citealt{brinkman01,guedel01}).

Finally, the X-ray luminosity $L_{\rm X}$ was determined in the energy range between 0.3 and 10 keV 
from the best-fit model, assuming a distance of 140 pc (e.g., \citealt{loinard05}, \citealt{kenyon94}).


\section{Results}\label{results}

\subsection{Light curves}\label{lightcurves}

\begin{figure*}
\centering
  \centerline{\hbox{
 \includegraphics[width=0.45 \textwidth]{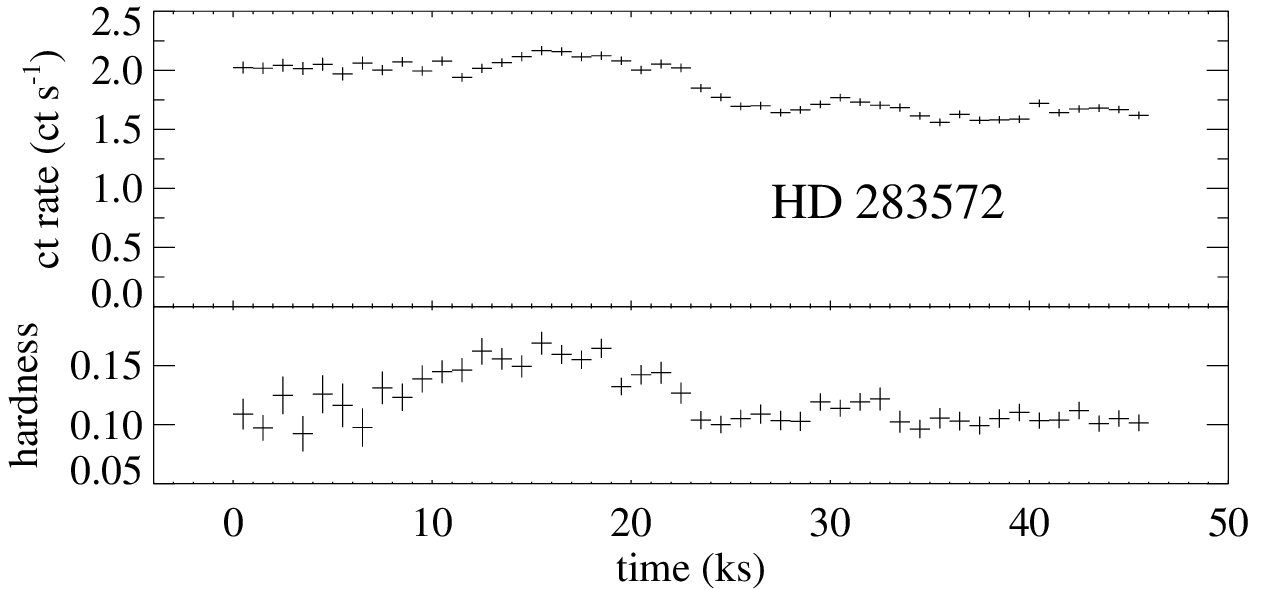}
 \includegraphics[width=0.45 \textwidth]{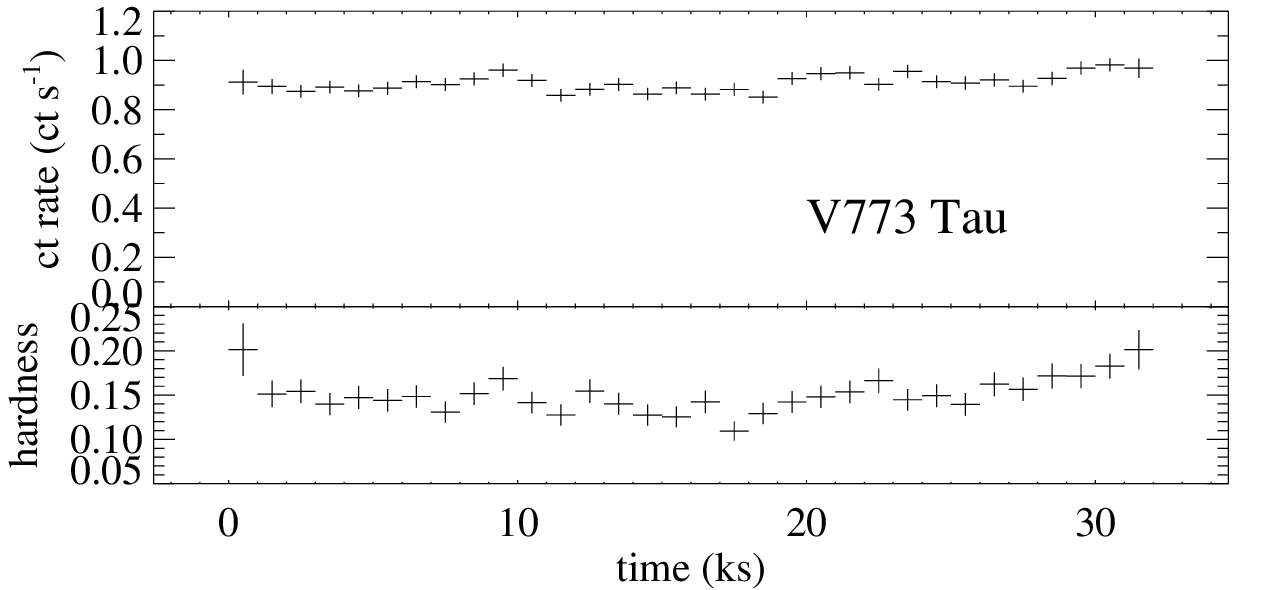}
    }}
  \centerline{\hbox{
 \includegraphics[width=0.45\textwidth]{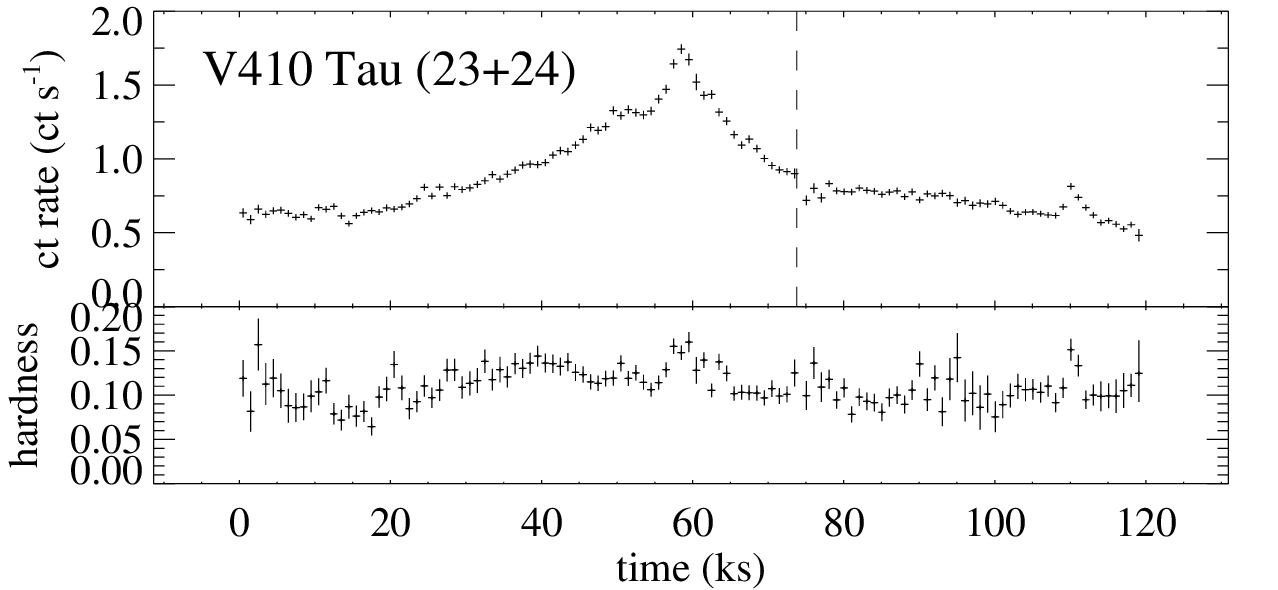}
 \includegraphics[width=0.45 \textwidth]{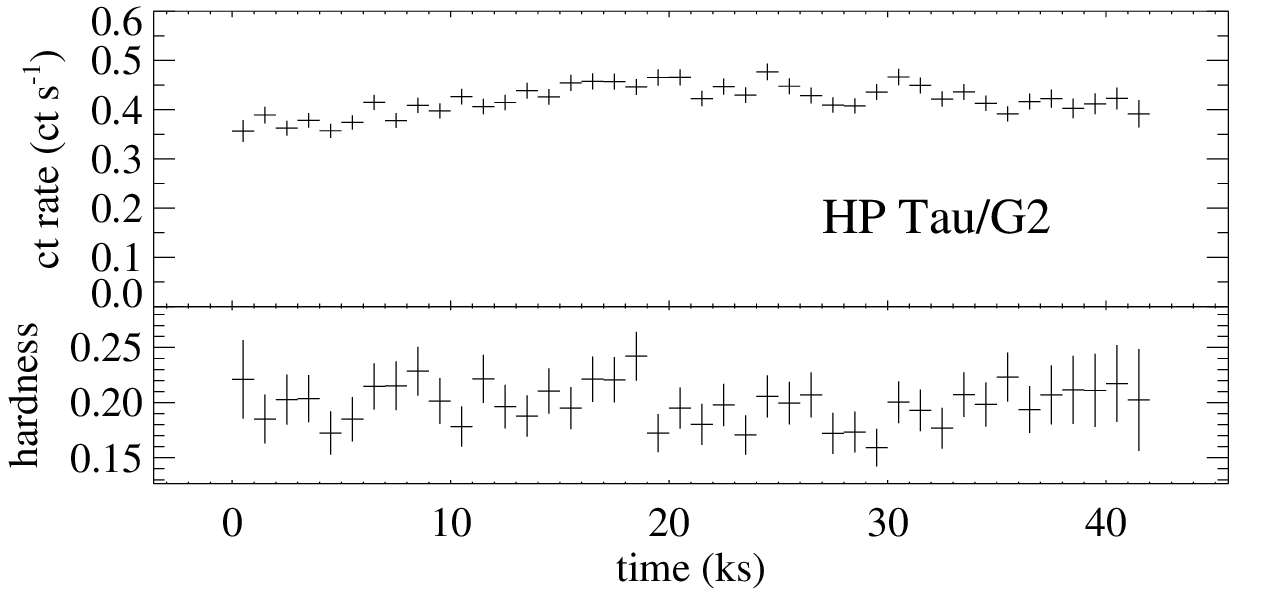}
    }}
  \centerline{\hbox{
 \includegraphics[width=0.45 \textwidth]{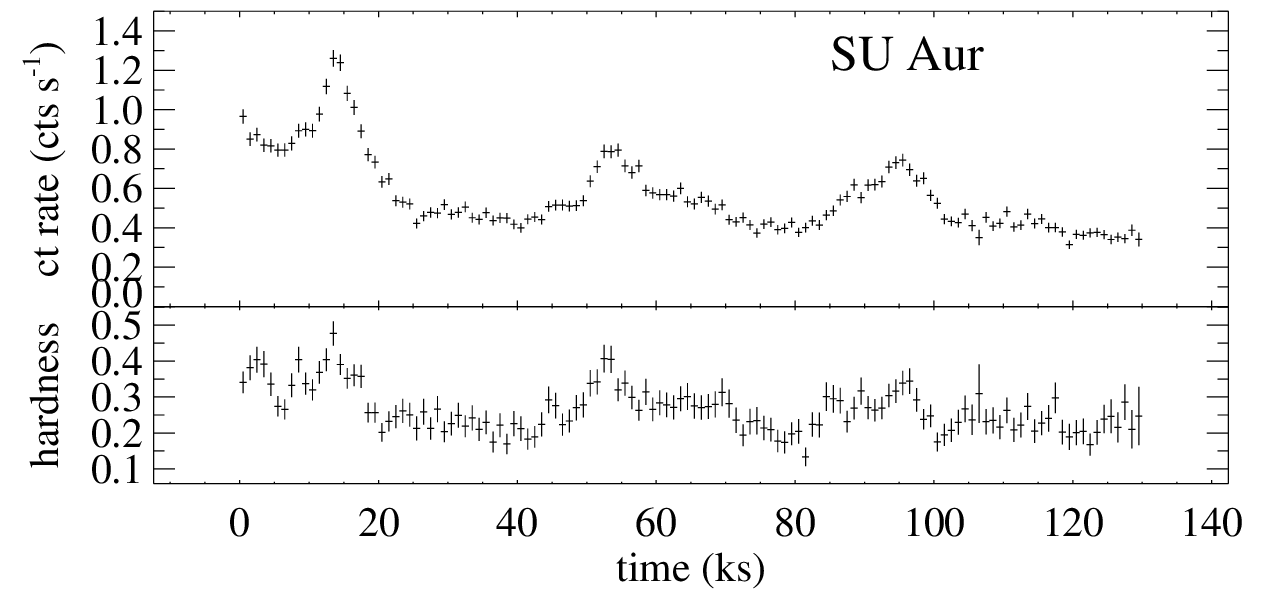}
 \includegraphics[width=0.45 \textwidth]{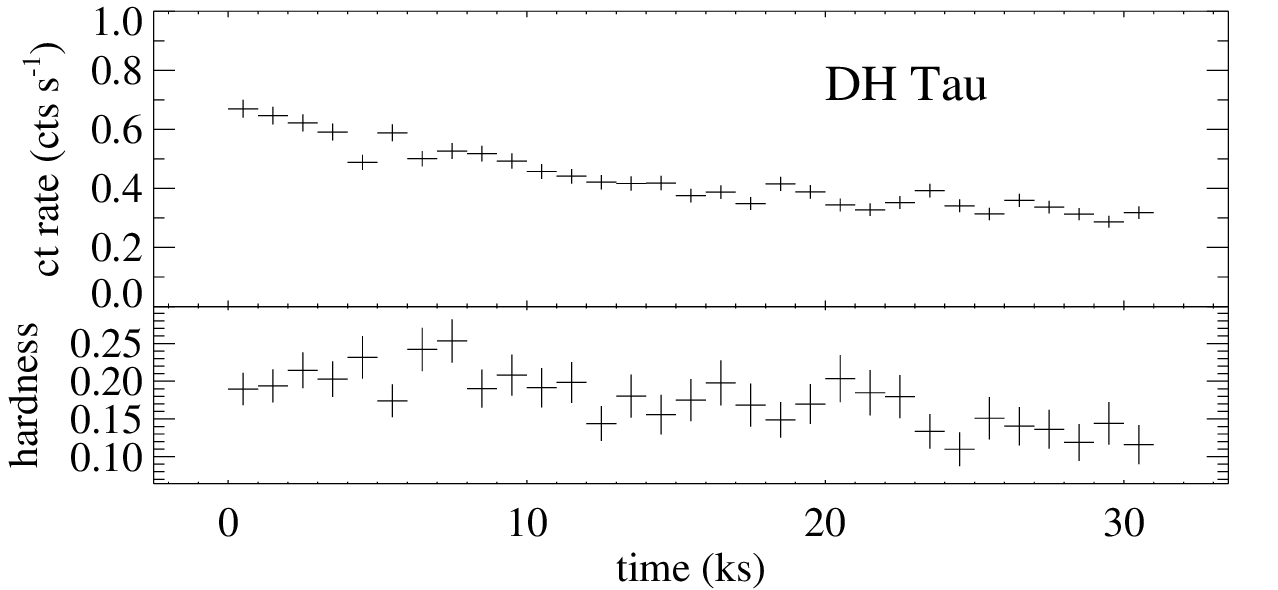}
    }}
  \centerline{\hbox{
 \includegraphics[width=0.45 \textwidth]{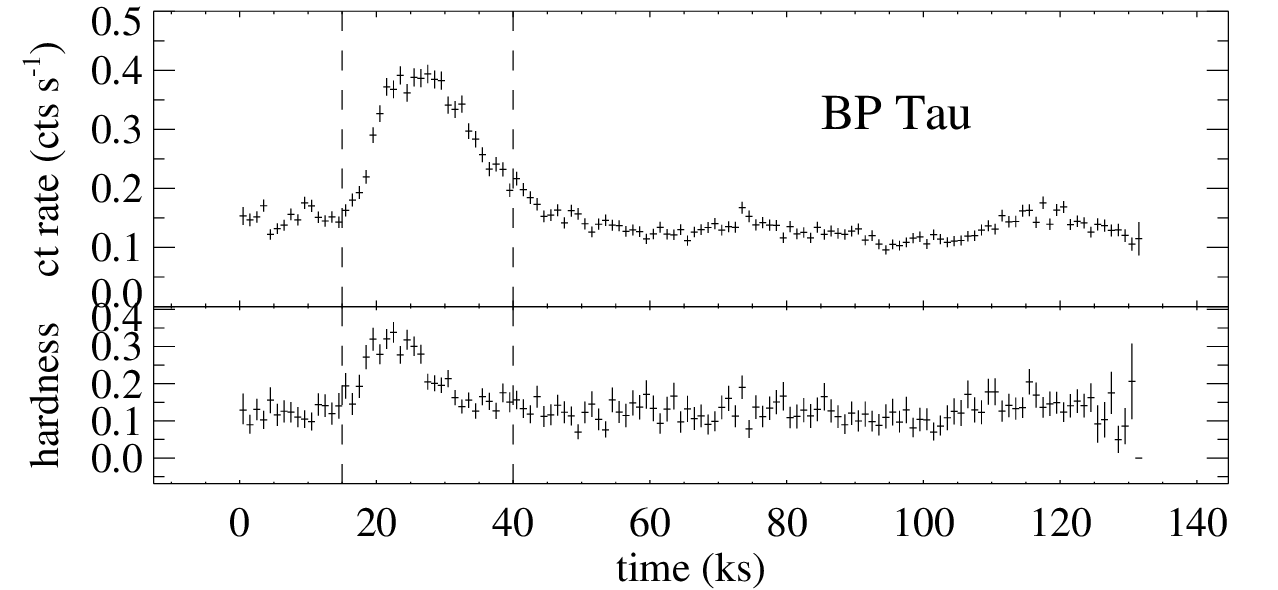}
 \includegraphics[width=0.45 \textwidth]{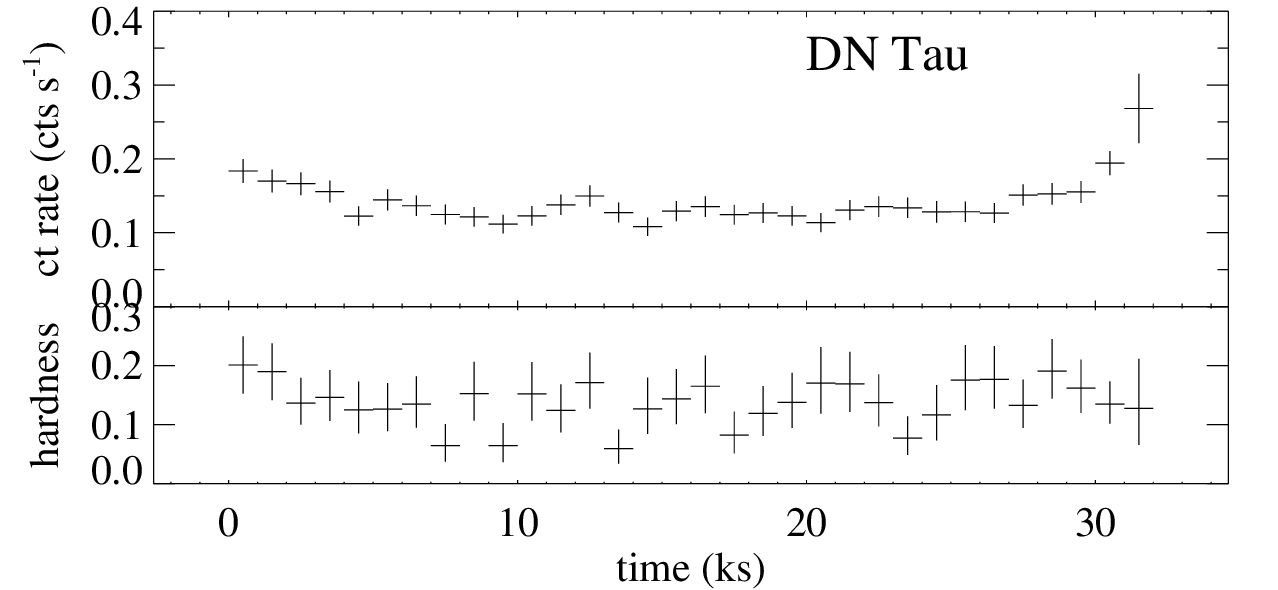}
    }}
  \includegraphics[width=0.45 \textwidth]{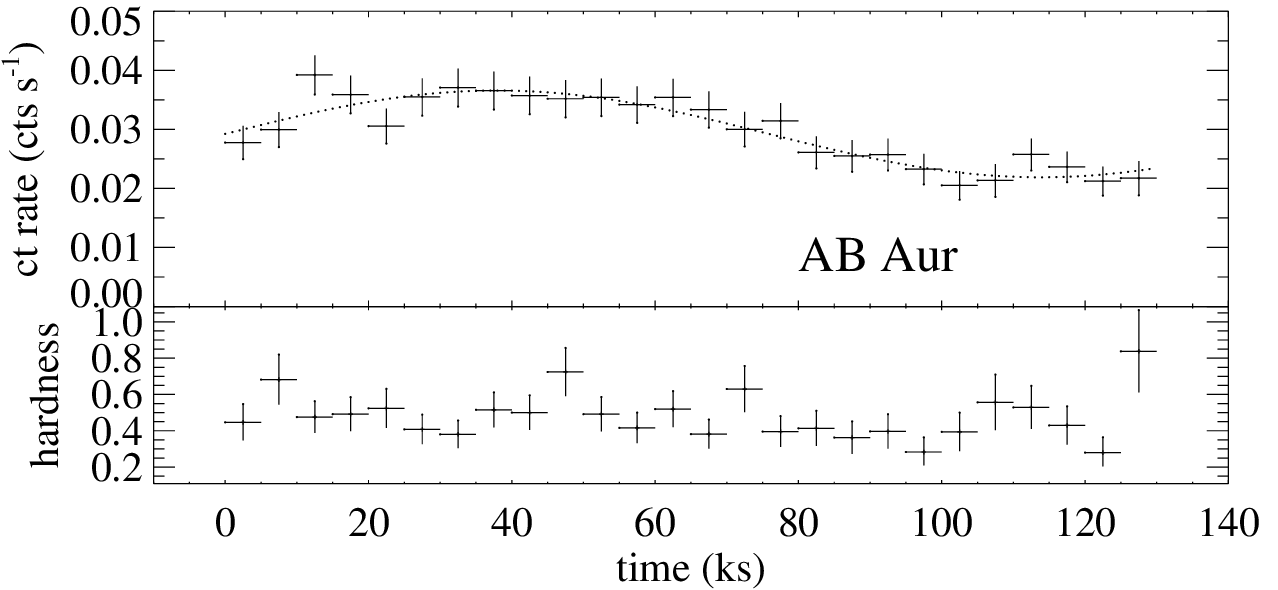}
  \caption{Light curves of the nine stars. Counts from all EPIC detectors
  have been combined, except for SU Aur and AB Aur, where data only from the two MOS detectors
  were available. Binning is to 1 ks, except for AB Aur, where 5 ks bins are used. Count rates
  refer to the energy range of 0.5-7.3 keV
  (except for AB Aur, where we used the range 0.3-4 keV). The size of the crosses in
  the x direction represent the bin width. Hardness is 
  given by the ratio between the hard band (1-4 keV for AB Aur, 2-7.3 keV for 
  all other stars) and the soft band (0.3-1 keV for AB Aur, 0.3-2 keV for all other stars). 
  The dashed line in the light curve of V410 Tau separates the observations of XEST-23 and XEST-24.
  The dashed lines in the BP Tau light curve
  mark the flaring time interval excluded from the spectral fit. The dotted curve overplotted on the
  AB Aur light curve shows a sinusoidal fit.}
  \label{fig_lc}
\end{figure*}

In Fig.~\ref{fig_lc}, the light curves of the nine stars are shown.
The star V410 Tau was observed twice consecutively (XEST-23 and XEST-24).
Both light curves are shown in our figure. For each light curve we also
show the hardness, defined as the ratio between the hard band counts
(1-4 keV for AB Aur, 2-7.3 keV for all other stars) and the soft 
band counts (0.3-1 keV for AB Aur, 0.3-2 keV for all other stars).

No strong variability (exceeding a factor of two
between minimum and maximum in count rate) is seen in 
the light curves of V773 Tau, HP Tau/G2, and DN Tau.
HD 283572 displays slow variability on timescales of 20-30 ks. 
The count rate varies by about 30\%.
Because the observation lasts about one third of the stellar rotation period,
the variation could be due to rotational modulation
of X-ray emitting regions in the corona, but equally well due to evolution
of active regions.

DH Tau's brightness decreases during the entire observation. 
Possibly, a large flare occurred
before the observation. Further evidence for this hypothesis 
is the continuously decreasing hardness.
Additional support  for a flare during the {\it XMM-Newton}
observation will be discussed in Sect.~\ref{lum}.

The light curve of the Herbig star AB Aur also shows variability.
We have fitted the curve with a sine function and we find
a modulation with a period of 42.2 hr. A similar modulation was also found in the
Mg\,{\sc ii} lines that are thought to be formed in the wind of AB Aur (see \citealt{telleschi06a}
for a detailed discussion).

The two observations of V410 Tau were taken consecutively. 
The light curve from XEST-23 (in the left panel in Fig.~\ref{fig_lc}) 
displays a flare, which has largely decayed in XEST-24 (in the right panel).
A small flare is visible at about 35 ks in
the second light curve. For the spectral fit, we  used only the second
observation (XEST-24). For a more detailed study of the X-ray variability of 
this and other TMC X-ray sources, see \citet{stelzer06} and \citet{franciosini06}.
Finally, flares are observed in the light curves of  SU Aur
and BP Tau.
SU Aur exhibits three
flares superposed on a slowly decaying light curve.
For BP Tau, the total contribution of the flare to the recorded counts
is modest, and the quiescent emission contained a sufficient number of counts so 
that we excluded the flare for the spectral fit. The time interval of the observation
ignored for the analysis is marked by the dotted lines (between 15 and 40 ks
after the observation start).

   \begin{figure*}
   \centering
   \includegraphics[width=\textwidth]{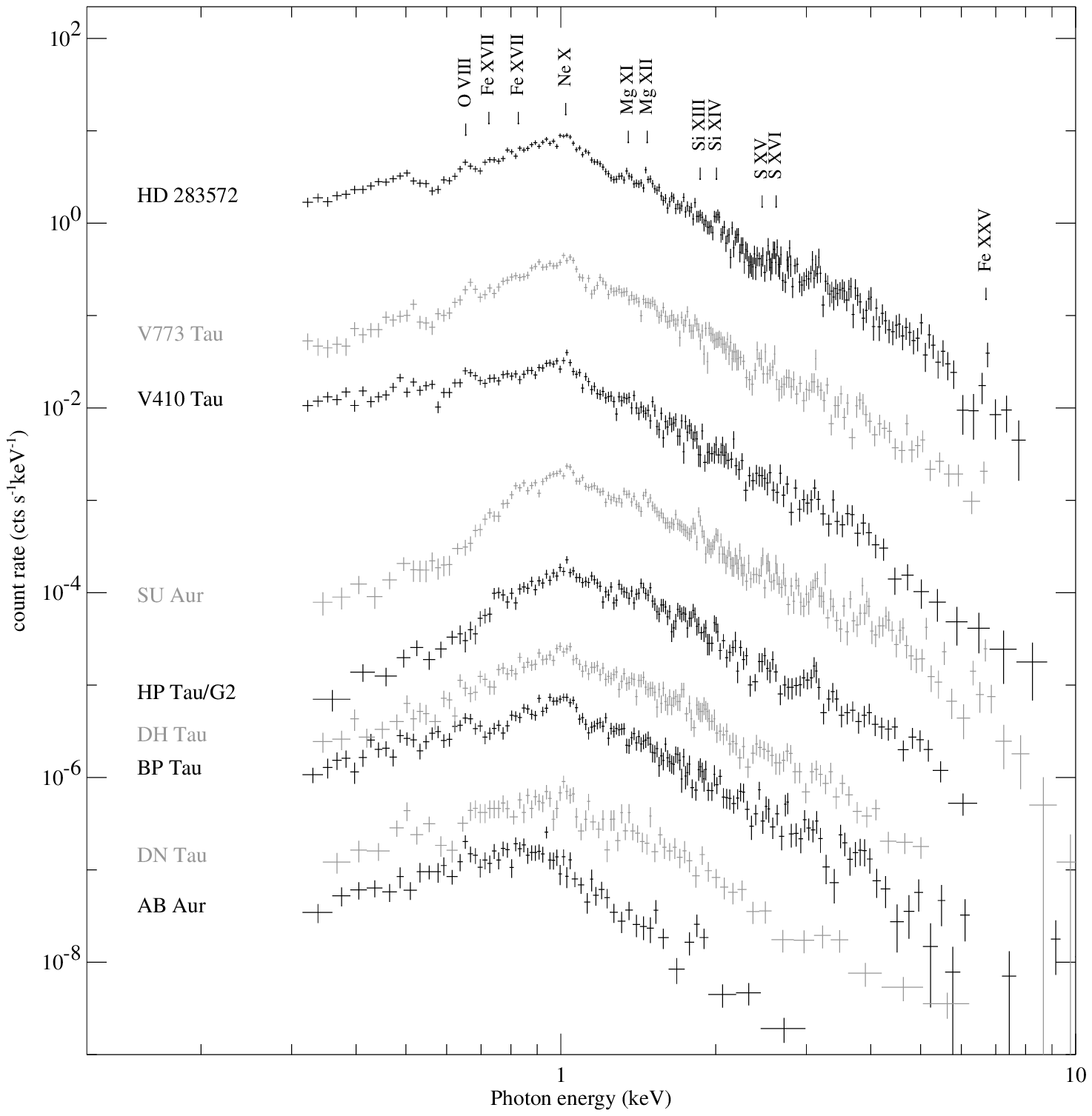}
      \caption{EPIC MOS1 spectra of the nine target stars. For plotting purposes
      the spectra have been multiplied with different factors: $10^{-5}$ for DN Tau, 
      $10^{-4}$ for BP Tau and DH Tau, $10^{-3}$ for HP Tau/G2,
      $10^{-2}$ for SU Aur, $10^{-1}$ for V410 Tau,
      1 for V773 Tau, and 10 for HD 283572.}
         \label{fig_epic_spec}
   \end{figure*}

\subsection{Spectra}

   \begin{figure*}
   \centering
   \includegraphics[width=\textwidth]{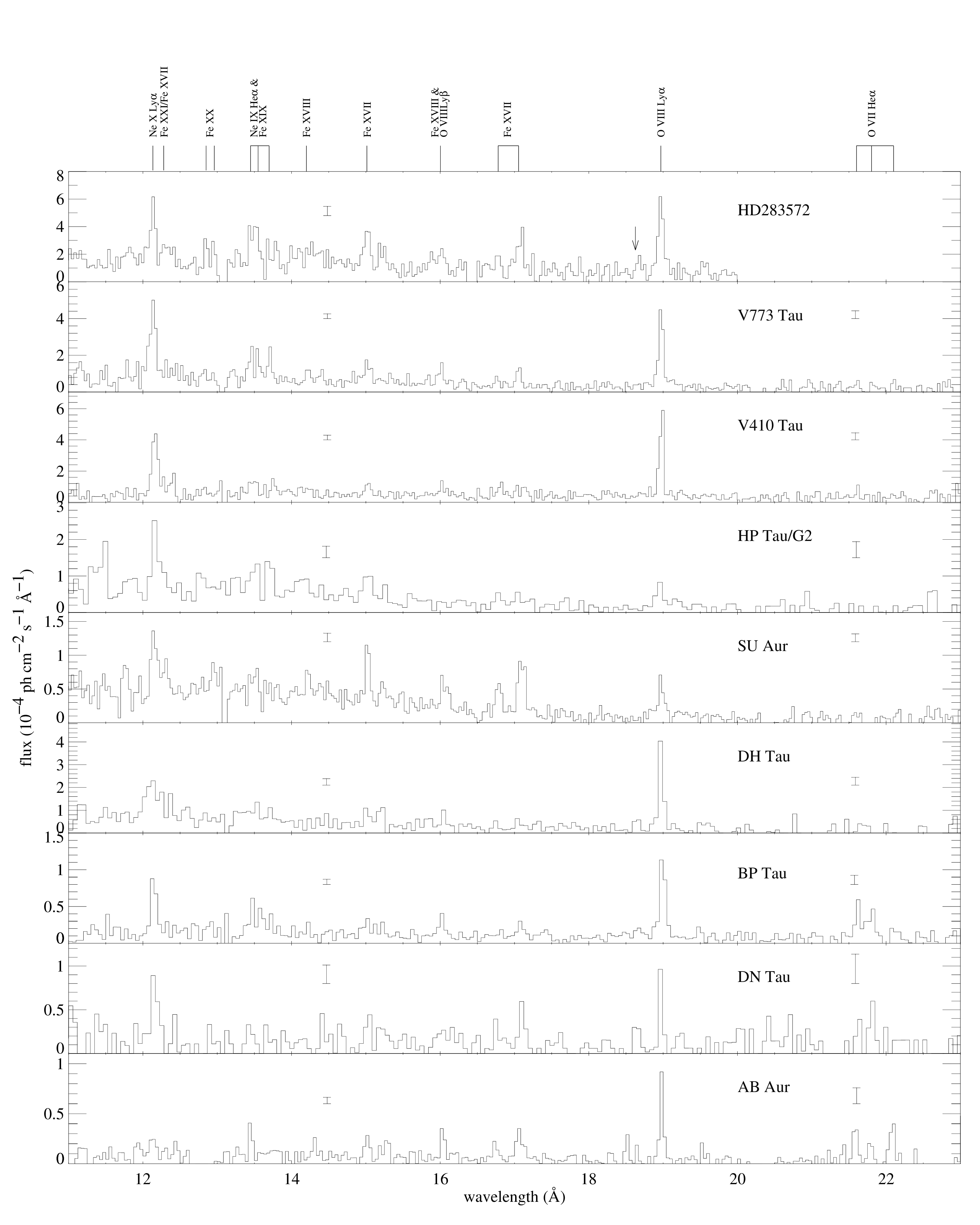}
      \caption{Co-added and fluxed spectra from the RGS1 and RGS2 instruments. 
               The spectra are background subtracted and are binned to a bin width of 0.066 \AA~for HP Tau/G2, 
               0.058 \AA~for DN Tau and DH Tau, 0.050 \AA~for BP Tau, 0.042 \AA~for AB Aur, and 0.035 \AA~for
               the other stars.  In each spectrum we also plot the typical 1$\sigma$ errors at 14.5 \AA\
               and at 21.6 \AA~(at the position of the O\,{\sc vii} line). The arrow in the HD 283572 spectrum 
               indicates the precise wavelength of the 
               O\,{\sc vii} He$\beta$ line at 18.63 \AA.}
         \label{fig_rgs_spec}
   \end{figure*}

The EPIC MOS spectra of the nine target stars are shown in Fig.~\ref{fig_epic_spec}.
For display purposes, we have multiplied the spectra with different factors
(see figure caption).
The effect of absorption is clearly visible on the low-energy slopes of the spectra.
The latter are steep for SU Aur and HP Tau/G2, demonstrating high photoelectric absorption,
while they are shallow in the other sources.
The spectra of HD 283572, SU Aur, and marginally V773 Tau show the Fe line complex
at 6.7 keV, which is a signature of very hot coronal plasma. We also note the presence of 
Ly$\alpha$ and He-like lines of Mg, Si, and S for these stars.
In contrast, the spectrum of AB Aur falls off rapidly above 1 keV, suggesting a dominance of
cool plasma.

Fig.~\ref{fig_rgs_spec} shows the fluxed RGS spectra of our sample.
The spectra of HD 283572, SU Aur, and HP Tau/G2 reveal a
strong continuum clearly pointing to hot plasma.
The spectra of SU Aur and HD 283572 show unusually large
flux ratios between the strongest Fe lines and the Ne\,{\sc ix} or
Ne\,{\sc x} lines when compared with other spectra, suggesting a
higher relative Fe abundance in these two stars.

A rather strong O\,{\sc vii} triplet is seen in 
the spectra of BP Tau, DN Tau, and AB Aur.
In the other CTTS spectra, the triplet is not present, probably because
the spectra are more absorbed.
Despite the low $N_{\rm H}$ for WTTS, their spectra do not reveal
the O\,{\sc vii} lines.
\footnote{The RGS1 was not in use for the observation of HD 283572
and the RGS2 does not cover the O\,{\sc vii} triplet due to a CCD chip failure.}

\begin{figure*}
\centering
\includegraphics[width=\textwidth]{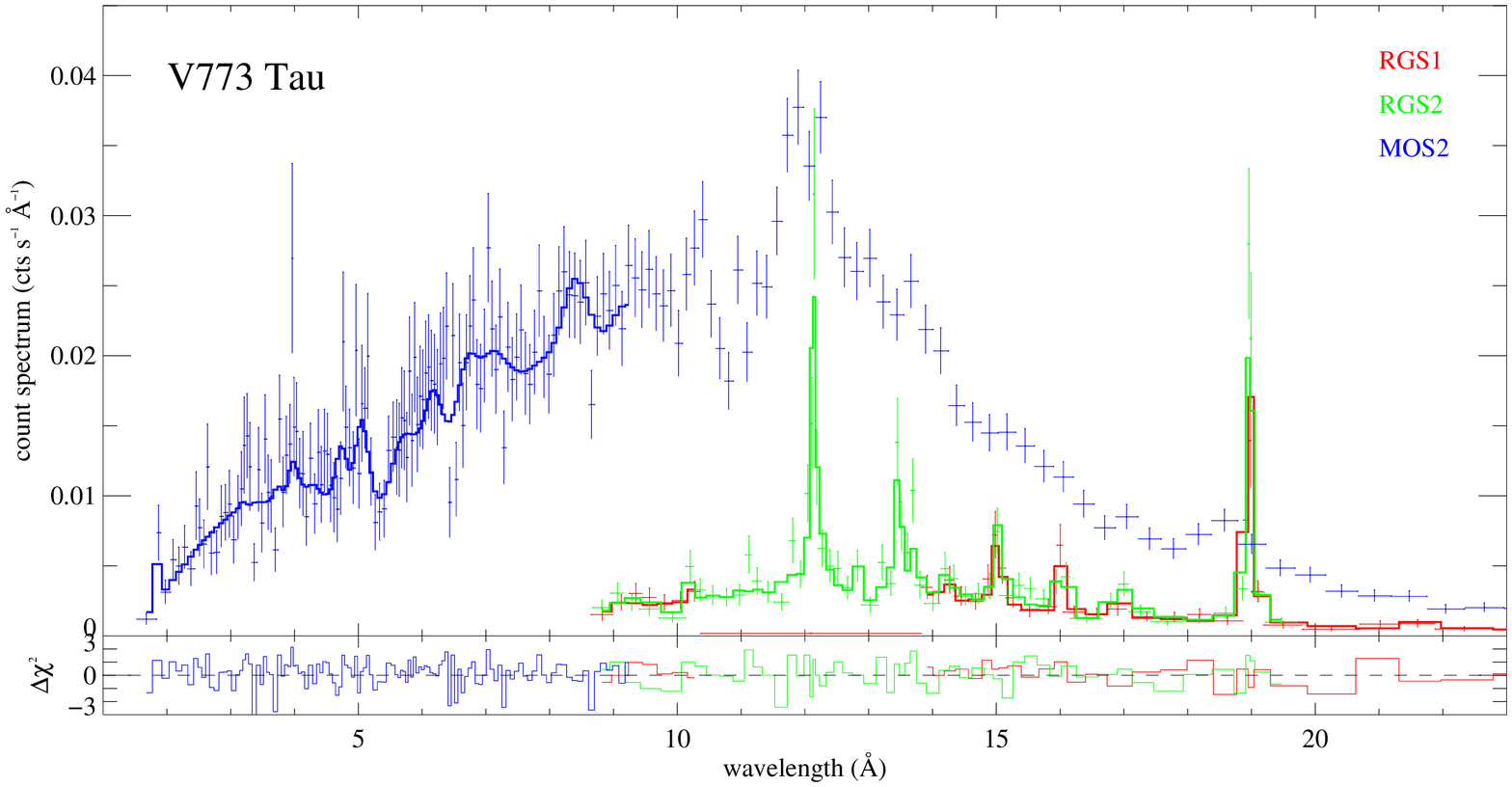}
   \caption{Data and best fit spectrum (EMD model) for V773 Tau.The best-fit model is shown by the histograms
      in the wavelength regions used for the fit.}
      \label{spec_fit}
\end{figure*}

\subsection{Thermal structure}

We now present results of our spectral fits. The fitted
parameters are listed in Table~\ref{tab:DEMfit} and
Table~\ref{tab:3Tfit} for the EMD model and the 2-$T$ or 3-$T$ model, respectively.
In Fig~\ref{spec_fit} we plot, as an example, the data and
the best fit of the EMD model for V773 Tau; the fits to the other spectra 
are similar.
The reduced $\chi^2$ are good for all fits ($\chi^2_{\rm red} \lesssim 1.3$) and are
very similar for the two approaches.

\begin{table*}
\caption{Results from the spectral fits using the EMD model.$^1$ }             
\label{tab:DEMfit}      
\centering                          
\begin{tabular}{l c c c c c}       
\hline\hline                 
Parameters                        & HD 283572 & V 773 Tau  & V 410 Tau &  HP Tau/G2 &  \\    
\hline                        
$N_{\rm H}$ [$10^{22}$ cm$^{-2}]$  & = 0.08~$^2$ & = 0.17~$^2$ & = 0.02~$^2$ &  = 0.41~$^2$ & \\
\hline
$T_0$ [MK]                   & 10.20 (9.20, 11.13)  &  8.72 (7.71, 9.74)  & 13.66 (10.79, 15.17)&  10.06 (9.07, 11.41)  & \\
EM [$10^{52}$ cm$^{-3}$]$^{3}$ & 127.98 & 75.18 & 45.22 &  90.06   &  \\
$\beta$                      &-0.84 (-0.99, -0.70) & -0.53 (-0.67, -0.40)  &  -1.59 (-1.80, -0.87)& -1.15 (-1.39, -0.99)  &   \\
\hline
O~$^4$    & 0.24 (0.19, 0.29)  & 0.48 (0.39, 0.59) & 0.39 (0.30, 0.47) & 0.20 (0.06, 0.38) & \\
Ne~$^4$   & 0.50 (0.40, 0.62)  & 1.30 (1.09, 1.55) & 1.08 (0.83, 1.31) & 0.65 (0.43, 0.91) &  \\
Mg~$^4$   & 0.51 (0.41, 0.63)  & 0.50 (0.37, 0.65) & 0.26 (0.15, 0.38) & 0.51 (0.35, 0.70) &  \\
Si~$^4$   & 0.25 (0.19, 0.32)  & 0.29 (0.21, 0.39) & 0.12 (0.04, 0.20) & 0.21 (0.13, 0.31) &  \\
S~$^4$    & 0.18 (0.08, 0.28)  & 0.64 (0.46, 0.83) & 0.48 (0.33, 0.65) & 0.26 (0.12, 0.41) & \\
Fe~$^4$   & 0.33 (0.27, 0.39)  & 0.29 (0.23, 0.36) & 0.20 (0.16, 0.24) & 0.43 (0.34, 0.56) &  \\
\hline
$T_{\rm av}$  [MK]           &   14.91 & 15.60  & 14.74 & 12.89 & \\
\hline
$L_{\rm X}$   [$10^{30}$ erg/s]$^5$ &  13.26 &  8.85 & 4.57 &9.60  &   \\
\hline
$\chi^2_{\rm red}$            &0.94 & 0.94& 1.24 & 0.92 &\\
dof                           & 251 & 208 & 173 & 131 & \\
\hline                                   
\hline                                   
Parameters                   & SU Aur   &  DH Tau    &  BP Tau  & DN Tau & AB Aur \\    
\hline                        
$N_{\rm H}$ [$10^{22}$ cm$^{-2}]$  & 0.32 & = 0.20~$^2$& = 0.06~$^2$ &  =0.07~$^2$ & 0.05\\
\hline
$T_0$ [MK]                   & 7.66 (6.95, 8.13) &  14.77 (12.37, 17.12) & 11.58 (10.03, 15.11) & 6.31 (4.16, 9.08) & 4.38 (2.69, 5.68)\\
EM [$10^{52}$ cm$^{-3}$]$^{3}$  & 57.84 &  87.88  &   11.24 & 15.37 & 5.12\\
$\beta$                       & -0.05 (-0.11, 0.06) & -2.34 (-3.00, -1.73) & -1.21 (-2.06, -0.93) & -0.66 (-1.06, -0.33) & -1.9 (-2.57,-1.52)\\
\hline
N~$^4$    & --                & --                & --                & --               & 0.57 (0.30, 1.13)\\
O~$^4$    & 0.30 (0.24, 0.38) & 0.49 (0.38, 0.64) & 0.46 (0.36, 0.59) &0.11 (0.05, 0.23) & 0.22 (0.13, 0.32)\\
Ne~$^4$   & 0.38 (0.28, 0.53) & 0.72 (0.55, 0.94) & 0.91 (0.66, 1.18) &0.45 (0.19, 0.84) & 0.62 (0.48, 1.04)\\
Mg~$^4$   & 1.17 (1.05, 1.43) & 0.33 (0.20, 0.48) & 0.50 (0.32, 0.69) &0.52 (0.27, 0.90) & 0.28 (0.13, 0.74)\\
Si~$^4$   & 0.64 (0.57, 0.79) & 0.16 (0.07, 0.25) & 0.26 (0.15, 0.38) &0.29 (0.12, 0.56) & 0.90 (0.60, 1.32)\\
S~$^4$    & 0.57 (0.45, 0.72) & 0.45 (0.30, 0.62) & 0.56 (0.34, 0.78) &0.11 (0.00, 0.56) & --\\
Fe~$^4$   & 0.67 (0.61, 0.77) & 0.16 (0.11, 0.23) & 0.18 (0.12, 0.24) &0.12 (0.06, 0.23) & 0.29 (0.22, 0.47)\\
\hline
$T_{\rm av}$  [MK]       & 20.07      &  13.76 & 14.28 & 11.30 & 4.71\\
\hline
$L_{\rm X}$   [$10^{30}$ erg/s]$^5$ & 7.79   & 8.20 & 1.16  & 1.24 & 0.39\\
\hline
$\chi^2_{\rm red}$        & 1.23     & 1.00 & 1.32 & 1.06 & 1.02\\
dof                       & 446     & 142 & 161 & 45 & 79\\
\hline                                   
\multicolumn{6}{l}{$^1$ 68\% error ranges are given in parentheses.}\\
\multicolumn{6}{l}{$^2$ Held fixed at values found in the XEST survey \citep{guedel06b}.}\\
\multicolumn{6}{l}{$^3$ Total EM integrated over temperature bins between $\log T=6-7.9$ [K] (see \citealt{guedel06b} for more details).}\\
\multicolumn{6}{l}{$^4$ Element abundances are with respect to solar values given by \citet{anders89} (\citealt{grevesse99} for Fe).}\\
\multicolumn{6}{l}{$^5$ Determined in the 0.3-10.0 keV band.}\\
\end{tabular}
\end{table*}

\begin{table*}
\caption{Results from the spectral fits using the 2T/3T model.$^1$ }             
\label{tab:3Tfit}      
\centering                          
\begin{tabular}{l c c c c c }        
\hline\hline                 
Parameters                         & HD 283572 & V 773 Tau  & V 410 Tau &  HP Tau/G2&      \\    
\hline                        
$N_{\rm H}$ [$10^{22}$ cm$^{-2}]$  & = 0.08~$^2$ & = 0.17~$^2$ & = 0.02~$^2$ & = 0.41~$^2$ &       \\
\hline
$T_1$ [MK]                   & 2.19 (1.33, 3.09)    & 4.51 (4.01, 5.77)    & 6.40 (3.96, 11.28)  &  8.86 (8.43, 9.19) &  \\
$T_2$ [MK]                   & 8.60 (8.33, 9.08)    & 9.15 (8.42, 10.93)   & 9.77 (5.64, 14.17)  &   25.72 (23.42, 28.37) & \\
$T_3$ [MK]                   & 26.03 (24.96, 27.12) & 29.39 (27.49, 32.46) & 24.78 (22.71, 26.69)&  --  & \\
EM$_1$ [$10^{52}$ cm$^{-3}$] & 18.77 (4.50, 47.13)  & 8.01 (4.29, 16.85)   & 8.61 (4.31, 17.16)  &   28.88 (20.95, 41.70) &  \\
EM$_2$ [$10^{52}$ cm$^{-3}$] & 30.33 (26.00, 35.88) & 20.28 (15.56, 22.43) & 11.25 (6.06, 18.78) &   35.41 (31.62, 40.19) &  \\
EM$_3$ [$10^{52}$ cm$^{-3}$] & 65.79 (62.50, 69.00) & 37.88 (34.30, 40.89) & 23.32 (20.61, 26.61) &  -- &  \\
\hline
O~$^3$    & 0.27 (0.19, 0.46) & 0.61 (0.47, 0.79) & 0.45 (0.37, 0.55) & 0.51 (0.20, 0.89) & \\
Ne~$^3$   & 0.91 (0.72, 1.12) & 1.60 (1.29, 2.01) & 1.10 (0.91, 1.35) & 1.10 (0.70, 1.56) & \\
Mg~$^3$   & 0.80 (0.65, 0.97) & 0.57 (0.42, 0.78) & 0.27 (0.15, 0.39) & 0.82 (0.56, 1.09) & \\
Si~$^3$   & 0.34 (0.25, 0.44) & 0.30 (0.19, 0.42) & 0.09 (0.00, 0.18) & 0.29 (0.17, 0.42) & \\
S~$^3$    & 0.22 (0.10, 0.35) & 0.57 (0.37, 0.78) & 0.48 (0.30, 0.66) & 0.27 (0.09, 0.46) & \\
Fe~$^3$    &0.59 (0.49, 0.68) & 0.37 (0.27, 0.49) & 0.18 (0.14, 0.23) & 0.72 (0.49, 0.97) & \\
\hline
$T_{\rm av}$  [MK]           & 12.96 & 16.38 &  14.84 & 15.94 & \\
\hline
$L_{\rm X}$   [$10^{30}$ erg/s]$^4$ &  13.56 &  8.75 & 4.60 & 9.26&  \\
\hline
$\chi^2_{\rm red}$            & 0.91 & 0.95 & 1.15 & 0.94 & \\
dof                           & 248 & 205 & 170 & 130 &  \\
\hline                                   
\hline                                   
Parameters                     & SU Aur     &  DH Tau    &  BP Tau  & DN Tau & AB Aur \\    
\hline                        
$N_{\rm H}$ [$10^{22}$ cm$^{-2}]$& 0.31 (0.30, 0.33) & = 0.20~$^2$ & = 0.06~$^2$ &  =0.07~$^2$ & =0.05~$^2$\\
\hline
$T_1$ [MK]                   &7.54 (7.30, 7.75)     & 4.66 (3.84, 6.08)    & 2.97 (2.69, 3.24) & 6.30 (5.36, 6.77) & 2.45 (2.10, 2.81)\\
$T_2$ [MK]                   &13.97 (12.05, 18.06)  & 11.28 (10.27, 12.32) & 9.42 (8.40, 10.83) & 26.09 (21.35, 31.66) & 6.99 (6.62, 7.41) \\
$T_3$ [MK]                   &38.14 (35.68, 41.14)  & 24.55 (21.36, 27.69) & 23.59 (21.58, 25.88) & -- & --\\
EM$_1$ [$10^{52}$ cm$^{-3}$] &11.27 (9.18, 14.04)   & 10.85 (6.05, 16.97)  & 2.18 (1.41, 3.28) & 4.57 (2.77, 9.50) & 2.11 (0.87, 3.67)\\
EM$_2$ [$10^{52}$ cm$^{-3}$] &7.25 (4.53, 14.35)    & 37.93 (24.16, 45.46) & 2.84 (1.99, 4.05) & 6.41 (5.28, 7.97)& 3.44 (3.06, 4.31)\\
EM$_3$ [$10^{52}$ cm$^{-3}$] &30.49 (24.66, 33.21)  & 34.79 (25.10, 46.39) & 6.17 (5.36, 7.05) & -- & -- \\
\hline
N~$^3$   & --                & --                & --                & --                & 0.45 (0.22, 0.78)   \\
O~$^3$   & 0.50 (0.40, 0.61) & 0.47 (0.32, 0.55) & 0.30 (0.23, 0.42) & 0.27 (0.10, 0.41) & 0.20 (0.16, 0.42) \\
Ne~$^3$  & 0.61 (0.46, 0.78) & 0.69 (0.51, 0.88) & 1.18 (0.83, 1.58) & 1.01 (0.44, 1.45) & 0.60 (0.41, 0.88) \\
Mg~$^3$  & 1.51 (1.33, 1.74) & 0.29 (0.15, 0.44) & 0.64 (0.42, 0.91) & 0.77 (0.39, 1.37) & 0.21 (0.09, 0.38)\\
Si~$^3$  & 0.79 (0.69, 0.92) & 0.13 (0.05, 0.23) & 0.30 (0.16, 0.45) & 0.41 (0.15, 0.79) & 0.70 (0.48, 1.01) \\
S~$^3$   & 0.64 (0.49, 0.79) & 0.44 (0.28, 0.62) & 0.48 (0.27, 0.72) & 0.00 (0.00, 0.31) & --        \\
Fe~$^3$  & 0.81 (0.73, 0.91) & 0.17 (0.11 ,0.23) & 0.27 (0.19, 0.35) & 0.21 (0.07, 0.37) & 0.23 (0.17, 0.32) \\
\hline
$T_{\rm av}$  [MK]      &   22.65   &  13.90 & 12.48 & 14.44 &  4.69 \\
\hline
$L_{\rm X}$   [$10^{30}$ erg/s]$^4$ & 7.40 &  8.35 & 1.20 & 1.19 &  0.40  \\
\hline
$\chi^2_{\rm red}$       & 1.20    & 1.00 & 1.26 & 0.99 & 1.06 \\
dof                       &  443    & 139 & 158 & 44 & 78  \\
\hline                                   
\multicolumn{6}{l}{$^1$ 68\% error ranges are given in parentheses.}\\
\multicolumn{6}{l}{$^2$ Held fixed at values found in the XEST survey \citep{guedel06b}.}\\
\multicolumn{6}{l}{$^3$ Element abundances are with respect to solar values given by \citet{anders89} (\citealt{grevesse99} for Fe).}\\
\multicolumn{6}{l}{$^4$ Determined in the 0.3-10.0 keV band.}\\
\end{tabular}
\end{table*}

For each star and each model we computed the average temperature ($T_{\rm av}$)
as the logarithmic average of  all temperatures used in the fit, applying
the emission measures as  weights.
We see a wide range of thermal properties, from the very hot
SU Aur with an average temperature of 20--23 MK, an EMD peaking at
about 7.7 MK and a nearly flat high-$T$  slope ($\beta = -0.05$),
to the cool AB Aur, with  $T_{\rm av} \approx $ 4.7 MK
and an EMD peaking at 4.4 MK with a steep high-$T$ slope ($\beta = -1.9$).
For HP Tau/G2, DN Tau, and AB Aur we fitted a 2-$T$ model instead
of the 3-$T$ model because a third component was not required for a good fit. 

Adding $\alpha$ as a free fit parameter did not change the abundances
significantly, but increased the error bars. The $\chi^2$ values were nearly
identical. The combination of fit parameters $\alpha$, $\beta$ and $T_{0}$
was ill-constrained for our stellar spectra. This is a consequence
of the weakness of our spectra, of the small number of lines available,
the breadth of the emissivity functions, and the interrelation
between emission measure and abundances.
We conclude that the EMD fit with fixed $\alpha$ is
satisfactory for our spectral analysis.

The average temperatures are similar for the two approaches.
The results of the EMD fit are also in quite good agreement with the results
from the EPIC fits (Table~\ref{epicfit}, after \citealt{guedel06b}).
Further, our results for SU Aur, BP Tau, and  HD 283572 are also within the error ranges 
of the results reported by \citet{robrade06} and \citet{scelsi05}.

\subsection{Abundances}

   \begin{figure*}
   \centering
   \includegraphics[width=0.44 \textwidth]{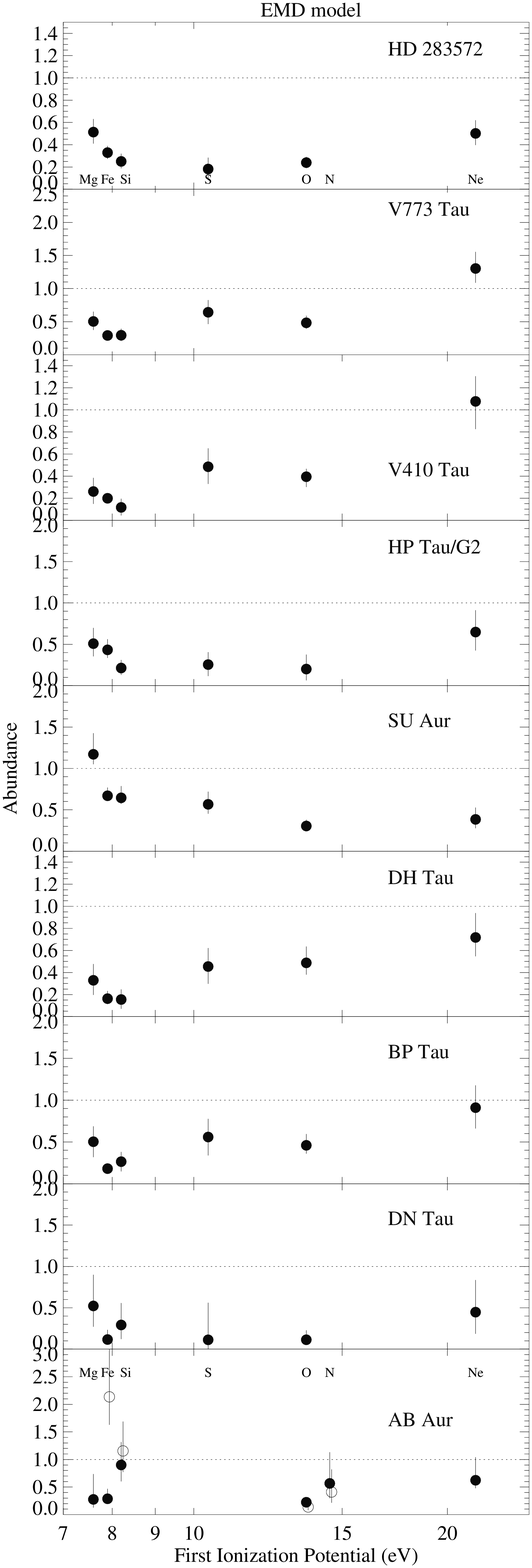}
   \includegraphics[width=0.44 \textwidth]{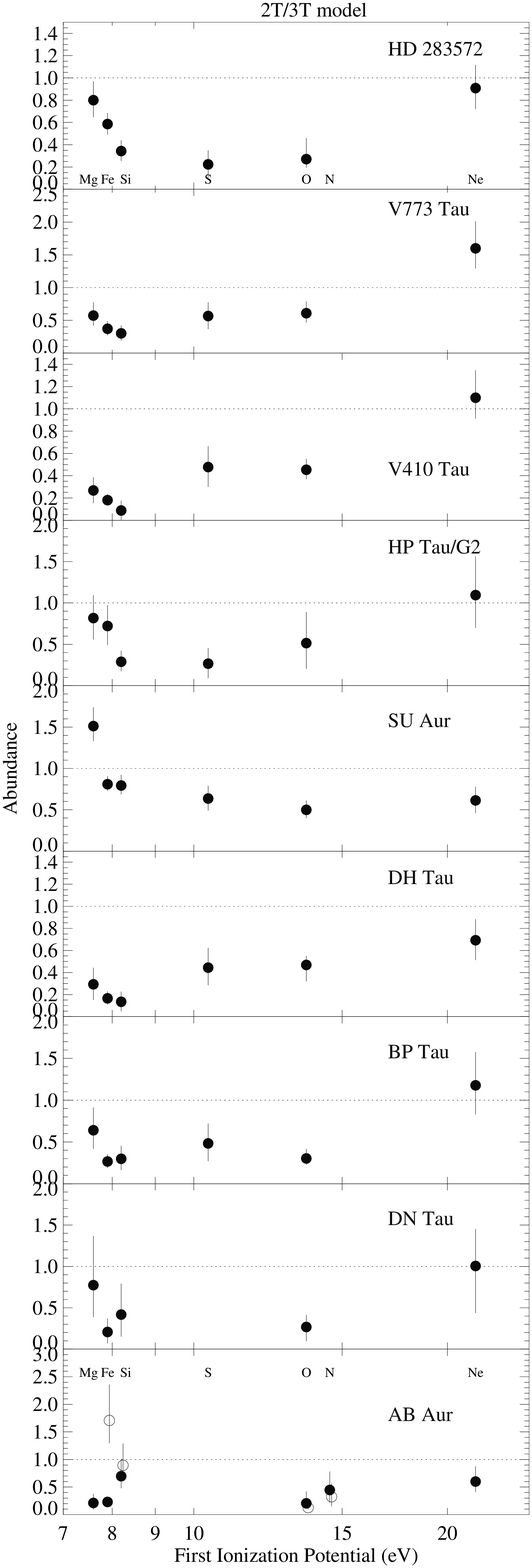}
      \caption{Element abundances
      normalized to solar photospheric values (\cite{anders89} and
      \cite{grevesse99} for Fe) derived from the EMD and 3T fits.
      Open circles for AB Aur: values normalized to the
      AB Aur photospheric abundances \citep{acke04}.}
         \label{fig_abun}
   \end{figure*}

The abundances are listed in Tables~\ref{tab:DEMfit} and
\ref{tab:3Tfit}, and are shown graphically in Fig.~\ref{fig_abun}
with respect to solar photospheric abundances.
For AB Aur the coronal abundances normalized to the AB Aur photospheric
values \citep{acke04} are also shown with the open circles.
The abundance patterns are similar for the two adopted models.
The abundances of HD 283572 first decrease
for increasing FIP, reaching a minimum around 10 eV (element S). 
For higher FIP, the abundances increase with FIP. A similar trend, although less
marked, can be observed for the abundances of HP Tau/G2.

The abundance patterns of V773 Tau, V410 Tau, DH Tau, and BP Tau are
consistent with an inverse FIP effect. Similar patterns are observed in
active stars and T Tauri stars \citep{telleschi05, argiroffi04}.
In contrast, the abundance pattern shown by SU Aur is peculiar among this sample: It is 
reminiscent of a solar-like FIP effect, i.e. elements with low FIP are more abundant than
elements with larger FIP.



As suggested above, we find that the Fe abundances are larger in
SU Aur, HP Tau/G2, and marginally in HD 283572 than in the rest 
of the sample. In the extreme case of SU Aur, the Fe abundance 
amounts to 0.67--0.81 times the solar photospheric value.
Such high Fe abundance values have been reported for
relatively inactive stars, while magnetically active stars
usually show a strong Fe depletion \citep{telleschi05,guedel04}.

The Ne/Fe abundance ratio reaches modest values for
SU Aur (0.6-0.75), HD 283572 and HP Tau/G2 (1.5),
and AB Aur (2.1-2.6). However, for the other T Tauri stars,
regardless of their accretion state, the Ne abundance is 4-6 times higher than the
Fe abundance. Such high Ne/Fe abundance ratios are unusual for main-sequence
stars (\citealt{telleschi05} reported Ne/Fe below 2 for their sample
of solar analogs), although they are reminiscent of ratios reported for
RS CVn binaries (Ne/Fe = 5.3--13.4, excluding Capella for which Ne/Fe=0.64, \citealt{audard03}).
A very large Ne/Fe abundance ratio has also been measured in TW Hya, reaching values
of 7--11 (\citealt{kastner02}; J. Kastner, 
private communication; \citealt{stelzer04}; B. Stelzer, private communication; referring to 
the solar photospheric abundances of \citealt{grevesse99} for Fe), in TWA 5 
\citep[$\approx$ 6,][]{argiroffi05} and in HD 98800
\citep[$\approx$ 4,][, J. Kastner, private communication]{kastner04}. For BP Tau, a 
high Ne abundance has been reported before \citep[$\approx$ 5,][]{robrade06}.

\subsection{X-ray luminosities}\label{lum}

Large changes in X-ray luminosity of coronal stars are usually accompanied 
by large changes of spectral parameters (e.g. temperature and densities),
in particular in the case of flares. Therefore, it is important  to
understand if the observed $L_{\rm X}$ is characteristic of the star
as judged from previous observations.
The X-ray luminosities from the two methods are in good agreement
for all stars.

For HD 283572 $L_{\rm X} \approx 1.3 \times 10^{31}$~erg s$^{-1}$
also agrees well with results found previously by \citet{favata98}
using other X-ray telescopes ($L_{\rm X} = [0.8-2.1] \times 10^{31} $~erg~s$^{-1}$),
and the value found by \citet[][$L_{\rm X} \approx 10^{31} $~erg~s$^{-1}$]{scelsi05}.

For V773 Tau we measured $L_{\rm X} \approx 8.8 \times 10^{30} $~erg~s$^{-1}$.
This value is slightly larger than that found by \citet{feigelson94} in their {\it ROSAT}
observation ($L_{\rm X}=  5.5 \times 10^{30} $~erg~s$^{-1}$; in the energy range 0.2-2 keV). 
Our $L_{\rm X}$ is consistent with the quiescent emission measured using {\it ASCA} by 
\citet{skinner97}: they  found $L_{\rm X} = 1.23 \times 10^{31}$~erg~s$^{-1}$ assuming a distance 
of 150 pc, corresponding to $L_{\rm X} = 1.07 \times 10^{31}$~erg~s$^{-1}$ at 140 pc.

Our $L_{\rm X}$ for V410 Tau ($L_{\rm X}=  4.6 \times 10^{30} $~erg~s$^{-1}$ for XEST-24) 
is smaller than the values found in the {\it ROSAT} observation by \citet{strom94} 
($L_{\rm X}=  1.3 \times 10^{31} $~erg~s$^{-1}$),
while it agrees well with the largest value reported by \citet{stelzer03} 
for a set of recent {\it Chandra} observations ($L_{\rm X}=  [3.2-4.0] \times 10^{30} $~erg~s$^{-1}$).
This comparison suggests that the flare seen in the first part of
our {\it XMM-Newton} observations has largely decayed in the second 
observation relevant for our study.

The $L_{\rm X}$ of SU Aur ($L_{\rm X}=  [7.4-7.8] \times 10^{30} 
$~erg~s$^{-1}$) is in agreement with the value measured by \citet{robrade06} 
for the same observation ($L_{\rm X}=  8.1 \times 10^{30} $~erg~s$^{-1}$), 
and with the value found by \citet{skinner98} ($L_{\rm X} = [8.4 \pm 0.09] \times 10^{30}$~erg~s$^{-1}$)
in the {\it ASCA} observation.
For BP Tau, our $L_{\rm X}$ ($\approx 1.2 \times 10^{30}$~erg~s$^{-1}$) is smaller than the value found by 
\citet[][$L_{\rm X} = 2.3 \times 10^{30}$~erg~s$^{-1}$]{robrade06}, probably because our $N_{\rm H}$ is smaller and we excluded the flare.

Overall, we find that the long-term variability illustrated by the above
comparison is compatible with the short-term variations seen in our light
curves. Variations within a factor of $\approx 2$ are common.

The DH Tau light curve is decreasing, suggesting that the source
is decaying after a strong flare. To test the activity level 
of this source, we compare our {\it XMM-Newton} observation 
with {\it ROSAT} observations reported in the {\sl roshri}
catalog in W3Browse.\footnote{a service of the Exploration of the
Universe Division at NASA/GSFC and the High Energy Astrophysics
Division of the Smithsonian Astrophysical Observatory (SAO),
http://heasarc.gsfc.nasa.gov/cgi-bin/W3Browse/w3browse.pl}
DH Tau was detected twice
by the {\it ROSAT} High-Resolution Imager (HRI) instrument, once
with a count rate of  0.020 ct~s$^{-1}$ as 1RXH J042941.3+263256
in observation rh202636, and once as a fainter source in the wings
of the nearby DI Tau (which was much fainter in the first observation)
in observation rh201088 (combined source 1RXH J042942.6+263250).
In this latter case, DH Tau was about four times  fainter than DI Tau,
and the total count rate of the two was 0.014 ct~s$^{-1}$.
We used PIMMS\footnote{http://heasarc.gsfc.nasa.gov/Tools/w3pimms.html}
to transform count rates to (unabsorbed) fluxes, adopting a temperature
of 10~MK and $N_{\rm H} = 2\times 10^{21}$~cm$^{-2}$ \citep{guedel06b}
to find $L_{\rm X} = 2.3 \times 10^{30}$~erg~s$^{-1}$ and $L_{\rm X} = 4 \times 
10^{29}$~erg~s$^{-1}$ in the two observations, i.e., 4-20 times
smaller than the $L_{\rm X}$ measured in our observation. We conclude
that the light curve is the result of the decay of a large flare 
starting prior to our observation.


\section{The He-like O\,{\sc vii} triplet}\label{triplet}


\begin{figure}
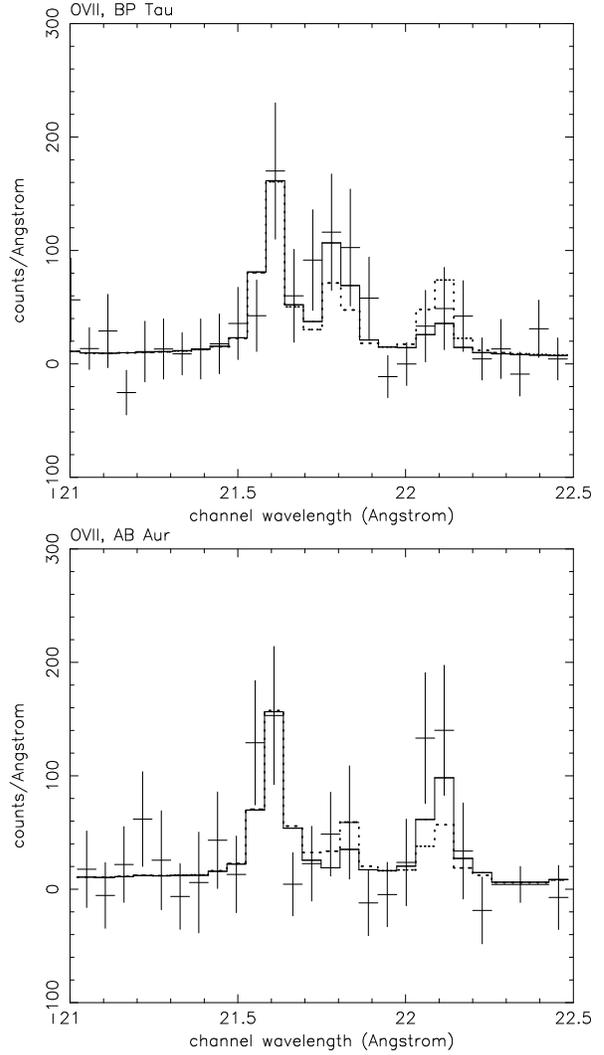

\centering
   \includegraphics[angle=-90, width=0.43\textwidth]{ovii_BP_bin5b_incl70c.ps}
    \includegraphics[angle=-90, width=0.43\textwidth]{ovii_AB_bin5c.ps}
     \caption{Fit of the O\,{\sc vii} triplets using variable electron density.
     Top panel, BP Tau: The best fit for $n_{\rm e} = 3.4 \times 10^{11}$~cm$^{-3}$ is
     plotted as a solid histogram, while the lower density limit
     ($n_{\rm e} = 6.4 \times 10^{10}$~cm$^{-3}$, 90 \% confidence) is
     shown by the dotted line. 
     Bottom panel, AB Aur: The solid histogram gives the best fit
    ($n_e \lesssim 10^{10}$~cm$^{-3}$), while the dotted histogram is
    for the 90\% upper limit to $n_e$ ($n_e < 1.3\times 10^{11}$~cm$^{-3}$).} 
     \label{fig_ovii}
\end{figure}

The flux ratio of the forbidden ($f$) line at
22.1 \AA~ and the intercombination ($i$) line at 21.8 \AA~
of O\,{\sc vii} is density-sensitive in the electron-density range
between $10^{10}$ cm$^{-3}$ and $10^{12}$ cm$^{-3}$
\citep{gabriel69}.
In the case of high densities, the more frequent collisions
trigger the excitation from the upper level of
the forbidden transition, $1s2s\ ^3S_1$, to the upper level of
the intercombination transitions, $1s2p\ ^3P_{1,2}$.
As a consequence,
the measured $f/i$ ratio becomes smaller.
Photoexcitation in an UV radiation field would also decrease
the $f/i$ ratio. The photon wavelength for the excitation
would correspond to the energy difference of the two upper
states, namely 1630 \AA~for the O\,{\sc vii} triplet. The UV radiation
field is thus important for stars with $T_{\rm eff} \gtrsim 10^{4}$~K,
i.e. only for AB Aur in our sample.

The measured ratio $\mathcal{R}=f/i$ of the forbidden to
intercombination line flux can be written as
\begin{equation}\label{foveri}
\mathcal{R} = {\mathcal{R}_0 \over 1 + \phi/\phi_c + n_e/N_c} = {f\over i}
\end{equation}
where $\mathcal{R}_0$ is the limiting flux ratio  at low densities and for O\,{\sc vii}
has a value of $\mathcal{R}_0 \approx 3.85$ \citep{blumenthal72}. $N_c$ is
the critical density at which $\mathcal{R}=\mathcal{R}_0/2$ ($N_c \approx 3.4
\times 10^{10}$~cm$^{-3}$ for O\,{\sc vii}, \citealt{blumenthal72}). The ratio $\phi/\phi_c$
is the radiation term needed for AB Aur.

In a thermal plasma, the flux of the resonance line, $r$, is larger than the flux
of $f$. Furthermore, for plasma with temperatures larger
than 1.5 MK, the sum $f+i$ is smaller than $r$, so that the ``G ratio''
$\mathcal{G}=(f+i)/r$ is smaller than unity \citep{porquet01}.
Given the low S/N ratio of our data in the wavelength range of interest,
it is difficult to fit the triplet lines individually and simultaneously fulfill the constraints
for the $\mathcal{R}$ and $\mathcal{G}$ ratios.
We instead proceeded as follows: we used the
best-fit results of the 3-$T$ model and then made use of the density-dependent 
calculations for the O\,{\sc vii} line fluxes as implemented in the XSPEC 
vmekal code. We kept all parameters fixed, except for
the electron density and the emission measure of the cooler component (to
allow for slight adjustments of the total O\,{\sc vii} line flux). Thus, the thermal structure
intrinsic to the model sets the correct requirement for $\mathcal{G}$, and
simultaneously fixes the continuum. 
The electron density $n_{\rm e}$ was then varied until a best fit for the fluxes
of the O\,{\sc vii} $i$ and $f$ lines was found. Only the wavelength
region of interest was used for the fit (between 21.4 and 22.3 \AA). The $\mathcal{R}$ ratio was finally
measured from the line fluxes.

In Fig.~\ref{fig_ovii} we present the O\,{\sc vii} triplet
for the stars BP Tau, and AB Aur, the only two
stars for which the triplet is clearly visible, together
with the best fits and upper or lower limits to the densities.
In the spectrum of DN Tau (Fig.~\ref{fig_rgs_spec}) an excess of counts is
present at the wavelengths of the O\,{\sc vii} triplet,
but the S/N ratio is too low to fit the triplet.

For BP Tau, the background was particularly high near
the O\,{\sc vii} triplet. We therefore decided to further
restrict the inclusion fraction of the cross-dispersion PSF to
70\% ($xpsfincl = 70$\%, see Sect.~\ref{observations}) specifically for
this wavelength region only. The resulting spectrum thus
contains fewer background counts and the triplet appears more
clearly, obviously at the cost of some source counts.
We also fitted the emission measure of the hottest component. 
This allows us to slightly adjust the continuum, that, because
of slight background subtraction inaccuracies, was not accurately represented.
We used bins of 56~m\AA~width for the fit.
The best fit is represented by a solid histogram in the top panel of 
Fig.~\ref{fig_ovii}. We find a best-fit electron density of $n_{\rm e}
= 3.4 \times 10^{11}$~cm$^{-3}$, corresponding to $\mathcal{R}= 0.23$.
The dotted line represents the 90\% lower
limit, corresponding to $n_{\rm e, min, 90} = 6.4 \times 10^{10}$~cm$^{-3}$,
and $\mathcal{R}= 1.07$. For the 68\% error we find
$n_{\rm e, min, 68} = 1.6 \times 10^{11}$~cm$^{-3}$, corresponding
to $\mathcal{R}=0.76 $. Given the low flux in the O\,{\sc vii} $f$ line,
we were unable to constrain upper limits to the densities.
The best-fit density is in agreement with the densities found
by \citet{schmitt05} and \citet{robrade06}, $n_{\rm e}
= 3 \times 10^{11}$~cm$^{-3}$ and $n_{\rm e}
= 3.2^{+3.5}_{-1.2} \times 10^{11}$~cm$^{-3}$, respectively.

In the bottom panel of Fig.~\ref{fig_ovii}, the O\,{\sc vii}
triplet of the Herbig star AB Aur is shown. Results on
the electron density measured in AB Aur have been discussed
in detail by \citet{telleschi06a}. Here, we will report
the main results in order to compare them with results from
CTTS.
We performed the fit again using a  bin width of 56~m\AA~
to find an electron density below the low-density limit,
$n_{\rm e} \la 10^{10}$~cm$^{-3}$ and $\mathcal{R} = \mathcal{R}_0$.
The dotted line in Fig.~\ref{fig_ovii}b corresponds to the 90\% upper limit, which is
$n_{\rm e, max, 90} \approx (1.3 \pm 0.4)\times 10^{11}$~cm$^{-3}$
and $f/i= 0.95$. For the 68\% upper limit we found
$n_{\rm e, max, 68} \approx (4.2 \pm 1.2)\times 10^{10}$~cm$^{-3}$,
corresponding to $f/i= 2.42$. 


\section{Discussion}\label{discussion}

\subsection{Abundance patterns}

The three stars with spectral type G (HD 283572, SU Aur, HP Tau/G2)
have similar properties. They were classified by \citet{herbig88}
as SU Aurigae-like stars, i.e., late type F to K stars showing an H$\alpha$ equivalent
width smaller than 10 \AA, weak Ca II emission, very broad absorption
lines ($v$sin$i > 50$ km~s$^{-1}$), and a relatively high optical luminosity.

\begin{figure}
\centering
   \includegraphics[width=0.48\textwidth]{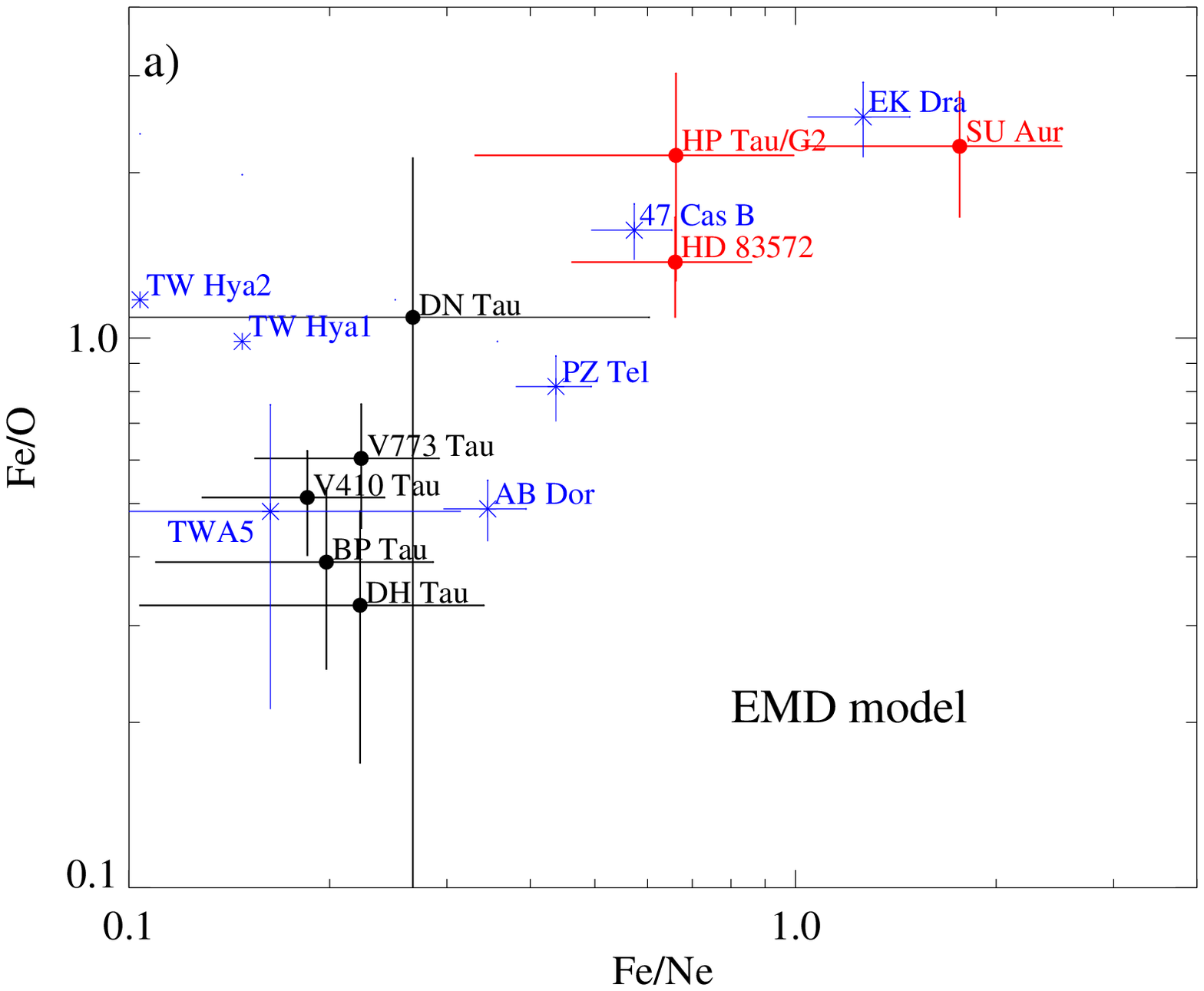}
    \includegraphics[width=0.48\textwidth]{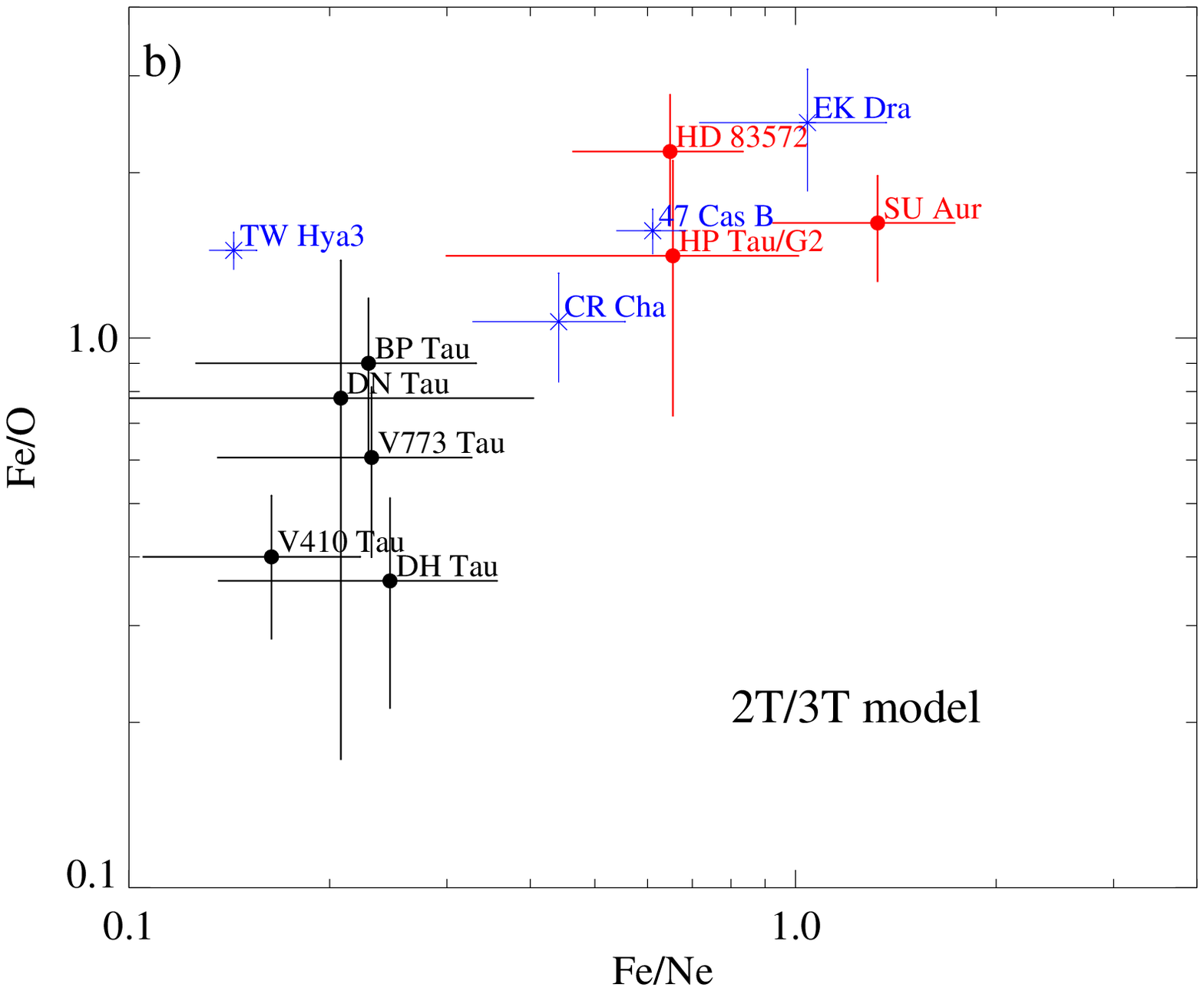}
     \caption{Fe/O abundance ratio as a function of the Fe/Ne ratio for abundances derived using
     an EMD model (a) and a 2/3-$T$ model (b). The K-M type stars 
     are plotted in black, while the G-type stars are in red. The ZAMS 47 Cas B and EK Dra
     are plotted in blue for comparison in both plots (from \citealt{telleschi05}). 
     The T Tau star TWA 5, the post T Tau star PZ Tel, the ZAMS AB Dor, TW Hya1 and 2 
     are plotted in (a) for comparison (the analysis were based
     based on an EMD method; see \citealt{argiroffi05,argiroffi04,garciaalvarez05,kastner02,stelzer04}).
     CR Cha and TW Hya3 have been added to (b) according to 3T fits reported 
     by \citet{robrade06}.}
     \label{fig_ratio}
\end{figure}

In X-rays, these stars display an abundance pattern with Fe
enhanced relative to oxygen (Fe/O $\approx$ 1.4 to 2.2), in 
contrast to the later-type stars in our sample that show an 
inverse FIP effect (Fe/O $\approx$ 0.35 to 1.0). Furthermore,
the Ne/Fe abundance ratio is relatively small for these three stars
(0.5--1.5) if compared with the other T Tauri stars in our sample
(Ne/Fe $\approx$ 4--6).
In Fig.~\ref{fig_ratio} $a, b$  we plot the Fe/O abundance ratio as
a function of the Fe/Ne ratio for the EMD and 2T/3T models,
respectively. The two stellar populations are different:
the K-M type stars (plotted in black) are lower in both
Fe/O and Fe/Ne than the G-type stars (plotted in red).
One possible explanation is that the abundances might be related
to mass or to surface gravity. However, then we would expect 
V410 Tau and V773 Tau to show abundance ratios similar to the
G-type stars, as their
masses are close to the masses of the SU Aurigae-like stars in our sample,
but this is not observed. \citet{scelsi05} have also
found similar abundances (when scaled to the Fe abundances) for three 
different G type stars with different masses at different evolutionary 
stages (HD 283572, EK Dra, and 31 Com). 
They concluded that surface gravity is not a determining factor for 
coronal abundance ratios.

In Fig.~\ref{fig_ratio} we also show for comparison
the abundance ratios of the TTS CR Cha \citep{robrade06},
TW Hya (labeled ``TW Hya1'' from \citealt{kastner02}, ``TW Hya2'' from \citealt{stelzer04}, 
and ``TW Hya3'' from \citealt{robrade06}), and TWA 5 \citep{argiroffi05}.
The abundances from these previous works have been converted to
the photospheric abundance normalization used here, i.e. \citet{grevesse99} for
Fe and \citet{anders89} for O and Ne.
CR Cha is of spectral type K2, TW Hya is K8, and TWA 5 is M1.5.
CR Cha and TW Hya3 were interpreted with a 3T model and are therefore
plotted in Fig.~\ref{fig_ratio}$b$, while TW Hya1, TW Hya2, and TWA 5 were interpreted with an
EMD method and are therefore plotted in Fig.~\ref{fig_ratio}$a$.
TWA 5 shows abundance ratios well compatible with our other K-type
TTS, while CR Cha's ratios are somewhat higher. We note, however,
that CR Cha is an early K star. 
TW Hya shows an Fe/Ne ratio similar to other K stars (especially in
the measurements from \citealt{kastner02} and \citealt{robrade06})
while its Fe/O ratio is high.

An abundance pattern similar to that of SU Aurigae-like stars 
was also observed in zero-age-main-sequence (ZAMS) stars with
spectral type G (47 Cas B, EK Dra, \citealt{telleschi05}).
For comparison, we therefore plot in Fig.~\ref{fig_ratio} $a, b$ the 
abundance ratios that were derived from a detailed EMD reconstruction 
and a multi-thermal fit, respectively, for these two G-type ZAMS stars. 
All G-type stars, independently of the evolutionary
stage, display similar abundance ratios.

Near zero-age main sequence K stars like
PZ Tel \citep{argiroffi04} and AB Dor 
\citep{guedel01,garciaalvarez05} show a classical inverse FIP pattern
similar to what we observe in our K-type T Tau stars.
The abundance ratios for PZ Tel \citep{argiroffi04} and AB Dor 
\citep{garciaalvarez05} are also shown in Fig.~\ref{fig_ratio} $a$.
Again, the abundance ratios are similar to the K-type TTS, although Fe/Ne
is somewhat higher.
We note, however, that both stars have somewhat earlier spectral types than 
typical for our sample
of T Tau stars: PZ Tel is classified as a K0 V star \citep{houk78}
and AB Dor is of spectral type K0-2 V \citep{vilhu87}. 

Further, we have checked the abundance ratios found
in active G-type and K-type main-sequence stars in the previous literature using
the compilation of \citet{guedel04} and a few recent references as given
in Table~\ref{tab:abratio}. For
these stars as well, both the Fe/Ne and Fe/O abundance ratios are larger for
G stars than for K stars. In Table~\ref{tab:abratio}
we summarize the abundance ratios for G and K-type stars.
The last three entries in Table~\ref{tab:abratio} are stars that
cannot be easily classified. AR Lac is composed of
a G and K star and both components contribute strongly to the X-ray emission \citep{siarkowski96}.
$\lambda$ And is a G8 giant star, i.e. intermediate between the two samples, in fact showing 
abundances similar to K-type stars. Finally, AB Aur is a Herbig star. 
These latter three stars are not included in the calculations of averages below.
We do not report the errors because we are mainly
interested in studying the distributions, i.e. their means and their standard deviations.
Further, the abundances reported in the table originate from different works based on different 
methods, implying that error estimates may not be consistent with each other.
Also, errors are not given in some papers.
The mean ratios for G stars are $\langle {\rm Fe/Ne} \rangle_{\rm G} = 1.02$ (standard 
deviation $\sigma=0.48$) and  $\langle {\rm Fe/O} \rangle_{\rm G} = 2.03$
($\sigma=0.42$). For K stars (using the mean of the three abundance ratios for TW Hya) we find $\langle {\rm Fe/Ne} \rangle_{\rm K}
 = 0.22$ ($\sigma=0.11$) and  $\langle {\rm Fe/O} \rangle_{\rm K} = 0.58$
($\sigma=0.32$), i.e. substantially lower than for G type stars.
Considering only the subsample of TTS, we find $\langle {\rm Fe/Ne} 
\rangle_{\rm G,TTS} = 1.03$ (standard deviation $\sigma=0.64$),  $\langle 
{\rm Fe/O} \rangle_{\rm G,TTS} = 2.92$ ($\sigma=0.47$), $\langle {\rm Fe/Ne} 
\rangle_{\rm K,TTS} = 0.23$ ($\sigma=0.10$), and  $\langle {\rm Fe/O} \rangle_{\rm K,TTS} 
= 0.71$ ($\sigma=0.35$), in agreement with the values found for the full sample.  
The Ne/O abundance ratios are also listed in Table~\ref{tab:abratio}.
We find  Ne/O to range between 1 and 3. \citet{drake05b} suggested that
this ratio might be sensitive to accretion; however, we find no difference in the Ne/O
ratio between CTTS ($\langle {\rm Ne/O} \rangle_{\rm C}=2.24$, $\sigma_{\rm C}=1.12$,
excluding TW Hya) and WTTS ($\langle {\rm Ne/O} \rangle_{\rm W}=2.75$,
$\sigma_{\rm C}=0.43$). The mean Ne/O ratio for all (K and G-type) stars in
Table~\ref{tab:abratio} (excluding TW Hya) is $\langle {\rm Ne/O} \rangle =2.76$
($\sigma =1.65$). The only star with exceptionally high Ne/O ratio 
remains TW Hya. This might be related to the older age of TW Hya and the
consequent evolution of grains in the disk \citep{drake05b}.
\citet{guenther06} recently reported an anomalously high Ne/O ratio
also for the CTTS binary V4046 Sgr.

\begin{table*}
\caption{Abundance ratios of active stars from this work and from the literature} 
\label{tab:abratio}      
\centering                          
\begin{tabular}{l c c c c c c}       
\hline\hline                 
Star & Type & Spec.Type & Fe/Ne & Fe/O & Ne/O & ref.\\ 
\hline
\multicolumn{5}{l}{Early to mid G-type stars}\\
\hline
HD 283572 & WTTS & G5       & 0.66 & 1.38 & 2.08 & this work\\
HP Tau/G2 & WTTS & G0       & 0.66 & 2.15 & 3.25 & this work\\
SU Aur    & CTTS & G2       & 1.76 & 2.23 & 1.27 & this work\\
47 Cas    & ZAMS& G0-2     & 0.57 & 1.57 & 2.74 & \citet{telleschi05}\\
EK Dra    & ZAMS& G0       & 1.26 & 2.52 & 2.0  & \citet{telleschi05}\\
Capella   & RS CVn& G1+G8  & 1.56 & 3.13 & 2.00 & \citet{audard03}\\
Capella   & RS CVn& G1+G8  & 1.41 & 1.65 & 1.16 & \citet{argiroffi03}\\
$\sigma^2$ CrB&RS CVn&G1+G8& 0.73 & 2.15 & 2.94 & \citet{suh05}\\
$\sigma^2$ CrB&RS CVn&G1+G8& 0.69 & 1.82 & 2.66 & \citet{osten03}\\
\hline
\multicolumn{3}{l}{Averages for G-type stars} & 1.02 & 2.03 & 2.25 & \\ 
\hline
\multicolumn{5}{l}{K-type stars (and early M for TTS)}\\
\hline
V773 Tau  & WTTS & K2       & 0.22 & 0.60 & 2.71 & this work\\
V410 Tau  & WTTS & K4       & 0.19 & 0.51 & 2.76 & this work\\
DH Tau    & CTTS & M1       & 0.22 & 0.33 & 1.47 & this work\\
BP Tau    & CTTS & K7       & 0.20 & 0.39 & 1.97 & this work\\
DN Tau    & CTTS & M0       & 0.27 & 1.09 & 4.09 & this work\\
CR Cha    & CTTS & K2       & 0.44 & 1.07 & 2.42 & \citet{robrade06}\\
TW Hya    & CTTS & K8       & 0.14 & 1.44 & 10.06& \citet{robrade06}\\
TW Hya    & CTTS & K8       & 0.15 & 1.00 & 6.67 & \citet{kastner02}\\
TW Hya    & CTTS & K8       & 0.10 & 1.16 & 11.60& \citet{stelzer04}\\
TWA 5     & WTTS? & M1.5       & 0.16 & 0.48 & 2.97 & \citet{argiroffi05} \\
AB Dor    & ZAMS & K0-2    &  0.34 & 0.49 & 1.42 & \citet{garciaalvarez05} \\
PZ Tel    & ZAMS& K0       & 0.44 & 0.81 & 1.87 & \citet{argiroffi04}\\
HR 1099   & RS CVn& K1+G5  & 0.15 & 0.36 & 2.40 & \citet{audard03}\\
UX Ari    & RS CVn& K0+G5  & 0.08 & 0.25 & 3.13 & \citet{audard03}\\
VY Ari    & RS CVn& K3-4   & 0.14 & 0.45 & 3.22 & \citet{audard03}\\
II Peg    & RS CVn& K2-3   & 0.07 & 0.14 & 2.02 & \citet{huenemoerder01}\\
V851 Cen  & RS CVn& K2     & 0.18 & 0.57 & 3.16 & \citet{sanzforcada04}\\
\hline
\multicolumn{3}{l}{Averages for K-type stars} & 0.22 & 0.58 & 3.00 & \\ 
\hline
\multicolumn{5}{l}{Other stars}\\
\hline
AB Aur    & Herbig & B9.5-A0  & 0.47 & 1.32 & 2.81 & this work\\
AR Lac    & RS CVn& K0+G2    & 0.46 & 1.23 & 2.68 & \citet{huenemoerder03}\\
$\lambda$ And&RS CVn& G8   & 0.19 & 0.57 & 3.00 & \citet{audard03}\\
\hline                                   
\multicolumn{7}{l}{$^1$ For TW Hya, the mean of the three abundance ratios (Fe/Ne resp. Fe/O) was used.}\\
\multicolumn{7}{l}{$^2$ Excluding TW Hya.}\\
\end{tabular}
\end{table*}

We conclude that a separation is visible between G-type stars
and mid-K-M-type stars, with G stars having a larger Fe/Ne
abundance ratio. Early K-type stars (like AB Dor, PZ Tel 
and CR Cha) show an intermediate Fe/Ne ratio.
A separation also exists in the Fe/O abundance ratio
if we exclude TW Hya. The latter star is however peculiar 
among TTS, since almost only cool plasma is present and the abundances 
refer, in contrast to other TTS, essentially to this cool plasma.  
It seems therefore that the abundance pattern in
the coronae of pre-main-sequence and near-ZAMS stars
relates to the spectral type, i.e. is a function of
the photospheric temperature. We caution that our sample 
is small, and further studies are needed to consolidate 
this trend.

\subsection{A soft excess in accreting stars}\label{softemission}

\begin{figure}
\centering
   \includegraphics[ width=0.48\textwidth]{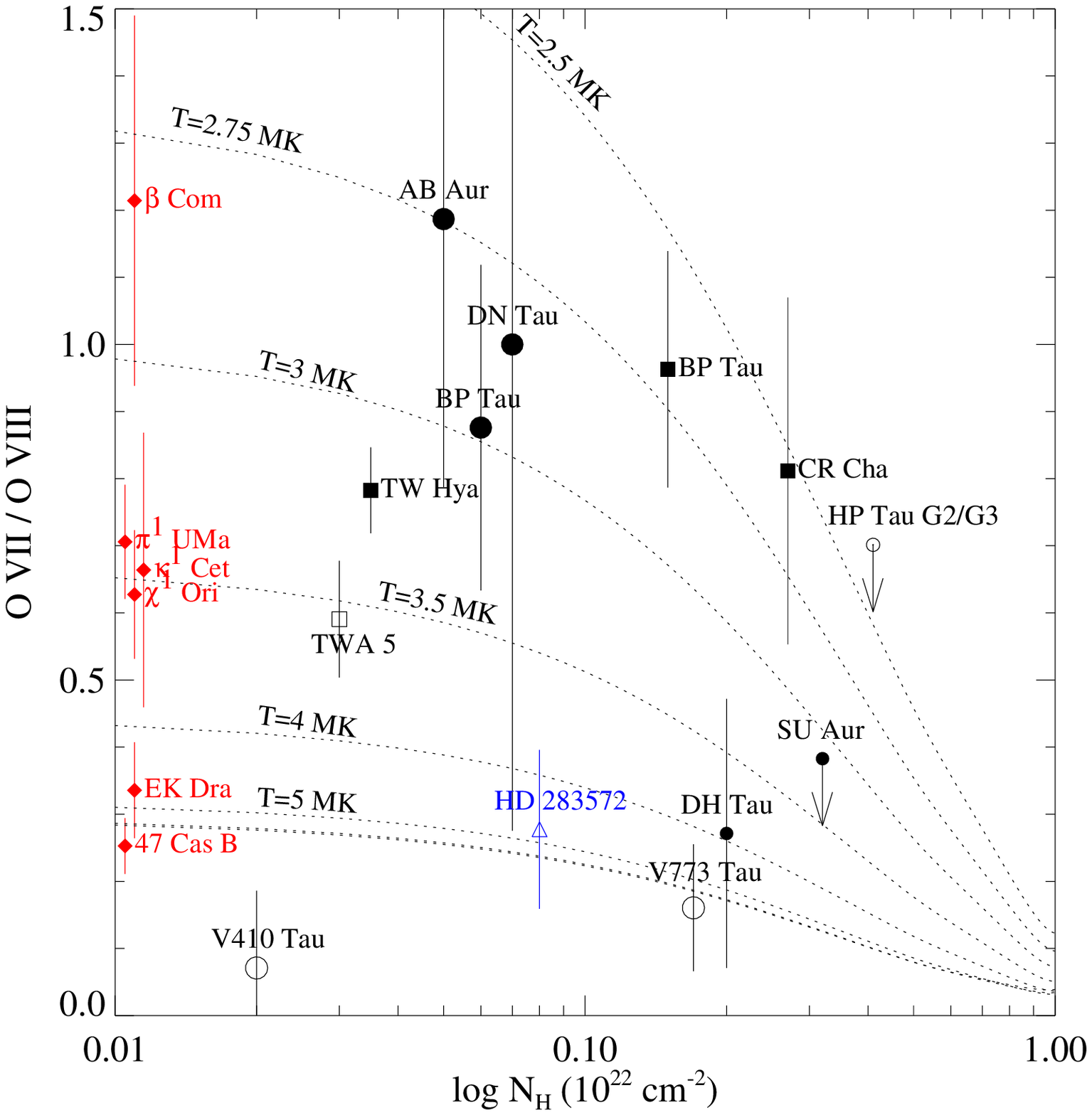}

     \caption{Ratio of the  fluxes measured in the O\,{\sc vii} triplet and the O\,{\sc viii}
     Ly$\alpha$ line as a function of $N_{\rm H}$. The sample of young stars
     presented in this paper is plotted with circles, while values taken from the
     literature are plotted with squares. The diamonds represent the main-sequence solar
     analogs of \citet{telleschi05}. HD 283572 is plotted with a triangle because
     the O\,{\sc vii} line counts are not measured but derived from the best fit model. Filled symbols
     denote accreting stars, open symbols are WTTS. Errors are 1 $\sigma$.}\label{o_ratio}
\end{figure}
In the RGS spectra of Fig.~\ref{fig_rgs_spec}, we
notice that the weakly absorbed spectra of accreting stars (BP Tau,
DN Tau, and AB Aur) display a relatively strong O\,{\sc vii} triplet
when compared with the O\,{\sc viii} Ly$\alpha$ line. On the contrary, in the spectrum
of V410 Tau (also subject to weak absorption, $N_{\rm H} = 2 \times 10^{20}$~cm$^{-2}$)
the O\,{\sc vii} triplet is not visible. This lets us hypothesize that a substantial cool
plasma component is present in the accreting stars but not in the 
WTTS.

The ratio between the fluxes in the O\,{\sc vii}  triplet and 
the O\,{\sc viii} line varies with temperature in the range of  $\approx 1-5$ MK.
In order to estimate this ratio, we derived the number of counts
at the wavelengths where the lines are formed (between 18.75-19.2 \AA\
for O\,{\sc viii} and 21.4-22.2 \AA\ for O\,{\sc vii}). From the
total number of counts measured in these wavelength intervals, we subtracted 
the number of background counts (scaled to the source area) and the number of counts
due to continuum (derived from the EMD best fit model). In the RGS1 the effective area drops 
substantially at wavelengths slightly shorter than O\,{\sc viii}, and
we therefore used the RGS2 spectrum to derive the number of counts
in this usually bright line (we used the RGS1
spectrum only for AB Aur, because in this spectrum the effective area of RGS2 drops
at wavelengths close to the O\,{\sc viii} line).
To obtain the fluence in the lines we 
divided the source counts by the effective areas at the relevant
wavelengths. 
The number of counts and the fluences derived with this method
are summarized in Table~\ref{tab:cts}. In the last column, we also
list the probability that the measured number of counts is due to 
Poissonian fluctuations in the background+continuum.
The RGS2 spectrum of DN Tau is difficult to interpret quantitatively
because it shows some excess in the background around the O\,{\sc viii}
line, resulting in too low flux in the line.
We therefore use the RGS1 to derive the the counts in the O\,{\sc viii} line. 

\begin{table*}
\caption{Number of counts and fluences for the O\,{\sc viii} and O\,{\sc vii} lines.
         Cts(tot), Cts(bkg), Cts(cont), and Cts(src) are the number of counts
         measured in the total spectrum, in the background spectrum, in the
         continuum (computed using the EMD fit results), and in the line 
         (Cts[tot]-Cts[bkg]-Cts[cont]). The last column give the probability that 
         the measured number of counts is due to fluctuations in the background and continuum.} 
\label{tab:cts}      
\centering                          
\begin{tabular}{l c c c c c c c c c}       
\hline\hline                 
Star & Line & Spectrum & Cts(tot) & Cts(bkg) & Cts(cont) & Cts(line) & Eff. Area & Fluence & Prob.\\ 
     &      &  used    &          &          &           &           & [cm$^2$]  & [ph/cm$^2$] & \\ 
\hline
BP Tau & O\,{\sc viii} & RGS2 & 81 & 10.57 & 11.53 & $58.90 \pm 9.13$ & 52.5 & $1.12 \pm 0.17$ & 0. \\
       & O\,{\sc vii}  & RGS1 & 80 & 26.26 & 13.44 & $40.30 \pm 9.24$ & 41   & $0.98 \pm 0.23$ & 1.2 10$^{-8}$ \\
V773 Tau&O\,{\sc viii} & RGS2 & 91 & 4.83  & 13.06 & $73.11 \pm 9.59$ & 53   & $1.38 \pm 0.18$ & 0. \\
        &O\,{\sc vii}  & RGS1 & 26 & 5.77  & 11.15 & $9.08 \pm 5.21 $ & 41   & $0.22 \pm 0.13$ & 2.4 10$^{-2} $\\
V410 Tau&O\,{\sc viii} & RGS2 & 82 & 3.61  & 14.78 & $63.61 \pm 9.10$ & 52   & $1.39 \pm 0.20$ & 0. \\
        &O\,{\sc vii}  & RGS1 & 40 & 11.45 & 24.52 & $4.03 \pm 6.51 $ & 41   & $0.10 \pm 0.16$ & 0.27 \\
HP Tau  &O\,{\sc viii} & RGS2 & 18 & 5.11  & 2.83  & $10.50 \pm 4.37$ & 53   & $0.20 \pm 0.08$ & 7.8 10$^{-4}$ \\
        &O\,{\sc vii}  & RGS1 & 6  & 7.61  & 1.11  & $< 2.4$          & 41   & $<0.06$         & 0.87 \\
SU Aur  &O\,{\sc viii} & RGS2 & 59 & 11.37 & 10.36 & $37.27 \pm 7.87$ & 53   & $0.90 \pm 0.19$ & 0. \\
        &O\,{\sc vii}  & RGS1 & 42 & 34.83 & 8.78  & $< 6.5$          & 41   & $<0.16$         & 0.62 \\
DH Tau  &O\,{\sc viii} & RGS2 & 67 & 6.59  & 10.02 & $50.39 \pm 8.27$ & 53   & $0.95 \pm 0.16$ & 0. \\
        &O\,{\sc vii}  & RGS1 & 44 & 25.39 & 8.03  & $10.58 \pm 7.62$ & 41   & $0.25 \pm 0.19$ & 4.5 10$^{-2} $\\
DN Tau  &O\,{\sc viii} & RGS1 & 17 & 3.37  & 3.06  & $10.57 \pm 4.21$ & 46   & $0.27 \pm 0.11$ & 3.8 10$^{-4} $\\
        &O\,{\sc vii}  & RGS1 & 20 & 6.66  & 4.32  & $9.02 \pm 4.62$  & 41   & $0.22 \pm 0.11$ & 9.1 10$^{-3} $ \\
AB Aur  &O\,{\sc viii} & RGS1 & 54 & 14.57 & 4.05  & $35.38 \pm 7.60$ & 43   & $0.82 \pm 0.18$ & 0. \\
        &O\,{\sc vii}  & RGS1 & 76 & 31.35 & 6.02  & $38.62 \pm 9.16$ & 38   & $0.98 \pm 0.25$ & 2.0 10$^{-8}$ \\
HD 283572&O\,{\sc viii}& RGS2 & $\approx 117$ & -- & 33.07 & $83.91 \pm 10.81$ & 52.5 & $1.60 \pm 0.21$ & 0. \\
         &O\,{\sc vii} & RGS1 & $\approx 55$  & -- & 36.51 & $18.18 \pm 7.40$  & 41   & $0.44 \pm 0.18$ & 2.6 10$^{-3}$ \\
\hline                                   
\end{tabular}
\end{table*}
 
The O\,{\sc vii}/O\,{\sc viii} fluence (or equivalently, flux) ratio is plotted in Fig.~\ref{o_ratio}
as a function of $N_{\rm H}$.
The dotted lines show the theoretical O\,{\sc vii}/O\,{\sc viii} 
ratio for isothermal plasma at a given temperature as a function of 
$N_{\rm H}$.
The  flux ratios measured in our sample 
are displayed by circles, while square symbols refer to measurements taken from the
literature (\citealt{robrade06} for TW Hya, BP Tau, and CR Cha; \citealt{argiroffi05}
for TWA 5). For HD 283572, because the RGS1 was not available and the RGS2
does not cover the O\,{\sc vii} wavelength region, we derived the  flux ratio from
the best-fit model.
Cool temperatures can be determined, apart from the O\,{\sc vii} He$\alpha$ line, 
by the O\,{\sc vii} He$\beta$ line at 18.63 \AA, but the line has not
significantly been detected. The feature that appears in the HD 283572 
spectrum in Fig.~\ref{fig_rgs_spec} is located slightly but significantly longward of the O\,{\sc vii}  
He$\beta$ line (at 18.75 \AA\ instead of 18.63 \AA) while the O\,{\sc viii} Ly$\alpha$ line
is located at its laboratory wavelength. The excess in flux
at  18.75 \AA\ in this spectrum is due essentially to a single
bin at 18.75 $\pm$ 0.03 \AA, 2$\sigma$ above the continuum, while the spectral fit
represents the data at the correct line wavelength accurately, implying a low 
flux also for the O\,{\sc vii} He$\alpha$ lines.
For a further check, we studied a spectrum of HD~283572 observed
by {\it Chandra} (Audard et al. 2007, in preparation). This spectrum
shows a well-developed O\,{\sc viii} line but no line at 18.62~\AA\ nor at
18.75~\AA. The 95\% confidence upper limit for the presence of excess flux
at 18.62~\AA\ is approximately 10\% of the O\,{\sc viii} Ly$\alpha$ flux.
Because the emissivity of the O\,{\sc vii} He$\beta$ line is about 14\% of
the emissivity of the O\,{\sc vii} He$\alpha$ $r$ line at 21.6~\AA\ and
the latter is about 60\% of the total triplet flux under typical conditions,
we conclude that a flux in the O\,{\sc vii} triplet must be lower than
O\,{\sc viii} Ly $\alpha$ line flux at the 95\% confidence level. In any
case, a strong line at 18.6-18.8~\AA\ as tentatively suggested by the
{\it XMM-Newton} RGS spectrum can be excluded.
The {\it Chandra} spectrum shows no indication of flux in the region of the
O\,{\sc vii} triplet, but the effective area of the HETGS instrument used
in this observation is too small to be useful for our study, because even
a triplet with a total flux equal to the flux in the O\,{\sc viii} Ly$\alpha$
line would not have been detected. 
For SU Aur and HP Tau/G2, the counts in the O\,{\sc vii} triplet were
very close to zero, and therefore only  95\% upper limits to the  flux
ratios are shown. We also note that these stars show the highest $N_{\rm H}$
in our sample, which is the reason for strong suppression of the O\,{\sc vii} 
triplet. Although the triplet is not explicitly visible 
in the spectra of V410 Tau, V773 Tau, and DH Tau, we measured a slight 
excess of counts in  the relevant wavelength interval. For these 
stars, we therefore plot their O\,{\sc vii}/O\,{\sc viii} flux
ratios at their best-fit loci.

For all WTTS, marked with open symbols in Fig.~\ref{o_ratio}, we
measure a low O\,{\sc vii}/O\,{\sc viii} ratio, even if $N_{\rm H}$
is small ($2 \times 10^{20}~{\rm cm}^{-2}$ for V410 Tau). The
temperatures corresponding to these line ratios ($T_{\rm oxy}$) are 
consistent with an isothermal plasma of $> 3.5$~MK.
The lack of strong O\,{\sc vii} is evident in Fig.~\ref{fig_rgs_spec}.
If the total flux in the O\,{\sc vii} were similar to the the flux in the 
O\,{\sc viii} Ly$\alpha$ line, then the O\,{\sc vii} $r$-line would be approximately 50-60\%
of the O\,{\sc viii} Ly$\alpha$ flux if $N_{\rm H}$ is low, which is not seen in 
Fig.~\ref{fig_rgs_spec} for these WTTS.  

For comparison, we also plot the O\,{\sc vii}/O\,{\sc viii} ratios of the six solar
analog stars presented by \citet{telleschi05}. These main sequence stars
are almost unabsorbed; for illustration purposes, we plot them at $N_{\rm H} \gtrsim 0.01
\times 10^{22}~{\rm cm}^{-2}$.
The O\,{\sc vii}/O\,{\sc viii}  flux ratios (or  the upper limits thereof)
that we measure in WTTS compare well with the same ratios measured in active
ZAMS stars such as 47 Cas B or EK Dra. With regard to the cool end of the
coronal emission measure distribution, WTTS and ZAMS stars seem to behave
similarly.

On the other hand, we measure a high O\,{\sc vii}/O\,{\sc viii} {\it only}
for accreting stars. For most of them,
the line ratio is consistent with $T_{\rm oxy} \approx 2.5-3$~MK.
These ratios are reminiscent of O\,{\sc vii}/O\,{\sc viii} 
found in rather inactive, evolved solar analogs. In contrast to
the CTTS, however, the coronae of those more evolved stars are {\it dominated}
by cool plasma, while much hotter plasma is common in CTTS.

For two CTTS, namely SU Aur and DH Tau, we measure an O\,{\sc vii}/O\,{\sc viii}
ratio lower than for the other accreting stars. However, we have noticed
that both stars are flaring (see Sect.~\ref{lightcurves}), which must
lower their O\,{\sc vii}/O\,{\sc viii} flux ratio.
The O\,{\sc viii} line is sensitive to the hot plasma, so that
its flux increases while the source is flaring. On the 
contrary, the O\,{\sc vii} triplet is insensitive
to the high temperatures measured in a strong flare. We
therefore expect the  O\,{\sc vii}/O\,{\sc viii} flux ratio
to decrease if the source is flaring.

$N_{\rm H}$ does alter the O\,{\sc vii}/O\,{\sc viii} ratio, but $N_{\rm H}$
cannot be made responsible for the lack of O\,{\sc vii} flux detected 
in some stars. The range of $N_{\rm H}$  measured in the WTTS sample
is in fact similar to the range measured in CTTS. 

According to the O\,{\sc vii}/O\,{\sc viii} ratio measured
in the sample investigated here, it is therefore possible that a 
soft excess is present in all accreting stars.
\citet{guedel06c} presented the spectrum of the CTTS T Tau. 
Although the corona of this star is extremly hot, a soft plasma 
component must be present
in order to explain the strong O\,{\sc vii} flux. 
The flux ratio measured in this star is  O\,{\sc vii}/O\,{\sc viii} =
$1.06 \pm 0.29$ for $N_{\rm H} = 0.48 \times 10^{22}$~cm$^{-2}$. The latter measurements confirm
the presence of a soft excess in T Tau, consistent with the
results that we find here. A specific discussion is presented 
by \citet{guedel06c}.

\begin{figure}
\centering
   \includegraphics[ width=0.48\textwidth]{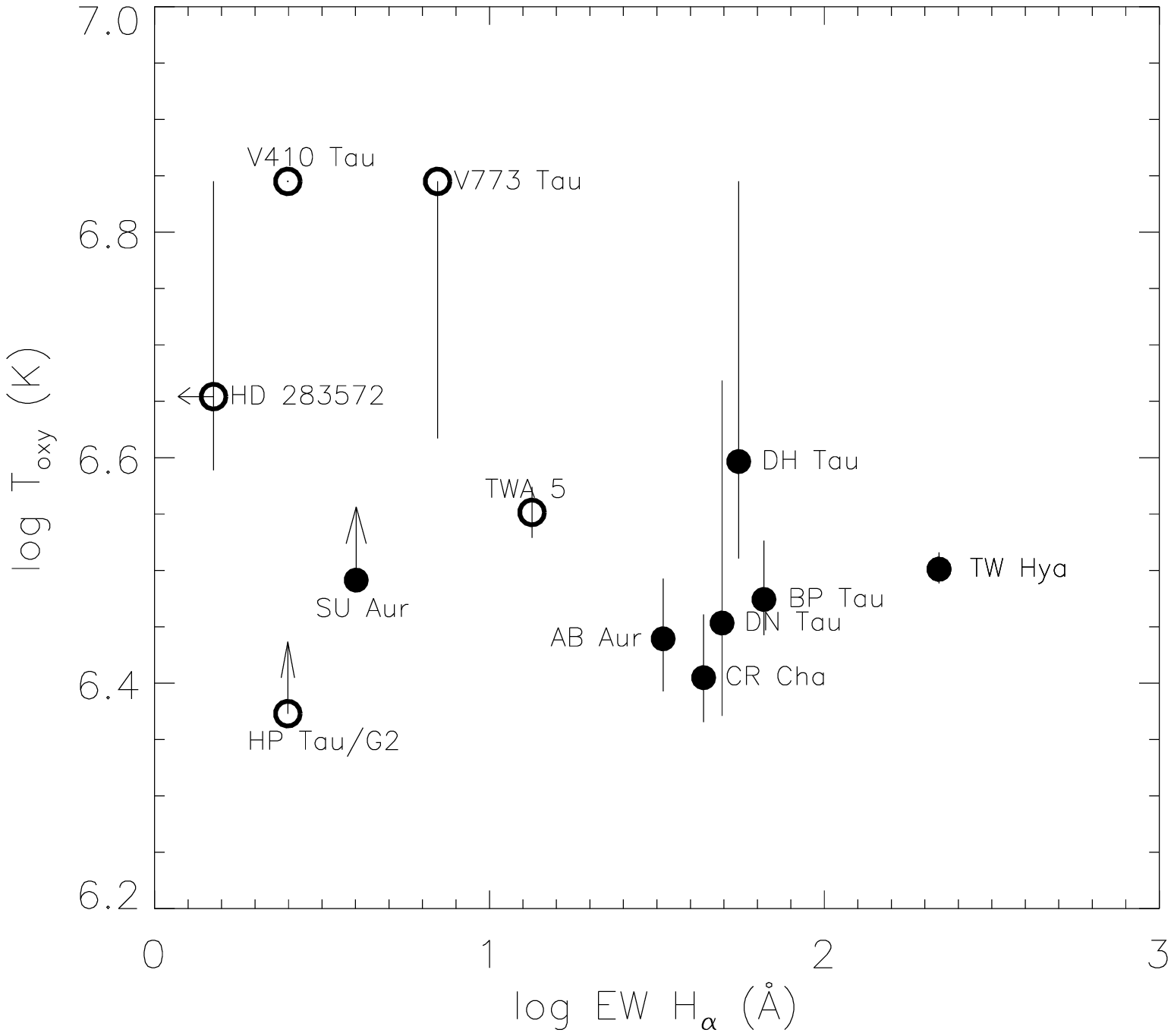}
   \includegraphics[ width=0.48\textwidth]{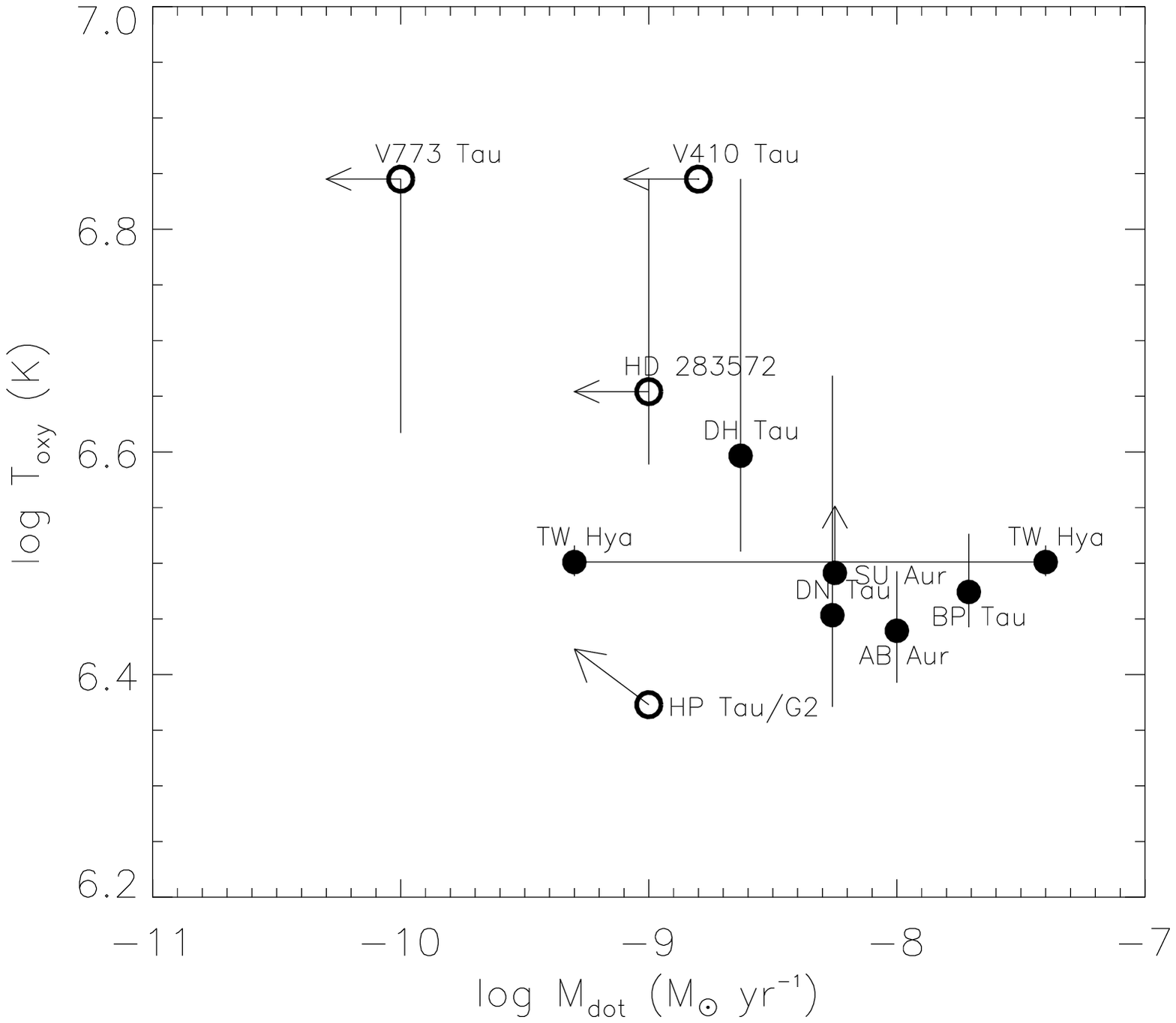}
     \caption{Temperature corresponding to the observed 
     O\,{\sc vii}/O\,{\sc viii} count ratio as a function 
     of H$\alpha$ equivalent width (top) and as a function 
     of the accretion rate $\dot{M}$ (bottom). Errors in $\log T_{\rm oxy}$
     are 1$\sigma$.}\label{corr}
\end{figure}

In order to illustrate the above trend using physical
properties that are not biased by other stellar
properties, we derive the temperature $T_{\rm oxy}$ 
using the loci for isothermal plasma plotted in 
Fig.~\ref{o_ratio}. $T_{\rm oxy}$ for each star
was computed using a spline interpolation. The
results are shown in Table~\ref{Toxy}.
$T_{\rm oxy}$ of V410 Tau and V773 Tau  are approximately or below 
the loci for 7 MK, above which the O\,{\sc vii}/O\,{\sc viii} ratio 
is no longer sensitive to temperature. We therefore assign 
a lower limit for  $T_{\rm oxy}$ of 7 MK to both stars.
In Fig.~\ref{corr}, we correlate the $T_{\rm oxy}$ with
stellar accretion parameters. 
 Filled and open circles represent the CTTS and WTTS, respectively.

\begin{table}
\caption{$T_{\rm oxy}$ derived from Fig.~\ref{o_ratio} for each star.} 
\label{Toxy}      
\centering                          
\begin{tabular}{l c }       
\hline\hline                 
Star & $T_{\rm oxy}$\\ 
\hline
BP Tau  &  2.97 (2.77 3.36)\\
V773 Tau&  7.0 (4.14 7.0)\\
V410 Tau&  7.0 (7.0 7.0)\\
HP Tau  &  $< 2.36$ \\
SU Aur  &  $< 3.10$ \\
DH Tau  &  3.95 (3.24 7.0)\\
DN Tau  &  2.84 (2.35 4.66)\\
AB Aur  &  2.75 (2.47 3.11)\\
HD 2835721& 4.51 (3.88 7.0)\\
\hline                                   
\end{tabular}
\end{table}

In the top panel of Fig.~\ref{corr}, we plot $T_{\rm oxy}$
as a function of the H$\alpha$ equivalent width. For the
stars in the XEST survey, the values of EW(H$\alpha$)
are taken from \citet{guedel06b} and references therein (see Table~\ref{prop}). The 
EW(H$\alpha$) values for TW Hya and TWA 5 are from
\citet{reid03}, namely 220 \AA~and 13.4 \AA, respectively, 
while the value for CR Cha is taken from \citet{gauvin92}, 
namely 43.6 \AA. For HD 283572, an equivalent width of
0 \AA~ has been reported \citep{kenyon98}; we plot an upper 
limit at EW (H$\alpha$) = 1.5 \AA~ for
illustration purposes. 
Generally, we find that stars with a large EW(H$\alpha$)
 show a low $T_{\rm oxy}$. One problem with using EW(H$\alpha$)
as an accretion signature is that for the same accretion rate ($\dot{M}$),
EW(H$\alpha$) is smaller in G-type stars than in K-type stars
due to formers' high continuum.

We therefore plot in the bottom panel of Fig.~\ref{corr} $T_{\rm oxy}$
as a function of $\dot{M}$ (Table~\ref{prop}, \citealt{guedel06b}). In the cases of HD 283572 
and HP Tau/G2, for which no upper limit for the accretion rate 
has been reported, we assign an upper limit of $\dot{M} < 10^{-9}$
~M$_{\odot}$ yr$^{-1}$ even if WTTS are thought not to be actively 
accreting at all. This upper limit was chosen according to the 
discussion in \citet{calvet04}: These authors find a well defined 
correlation between $M$ and $\dot{M}$. In the mass ranges between 
0.1 and 1 $M_{\odot}$, accreting stars could have been detected 
down to a limiting $\dot{M}$ two orders of magnitude lower than 
observed values. In the case of intermediate-mass TTS, the photospheres
are brighter than for lower mass stars, so that lines with 
similar flux have a smaller equivalent width. For stars with 
masses similar to HD 283572 and HP Tau/G2, accretion rates between 
$10^{-9}$ and $10^{-8}$~M$_{\odot}$ yr$^{-1}$ were clearly 
identified. 
For V773 Tau and V410 Tau we used upper limits for the accretion rate 
as reported in the XEST catalog (see \citealt{guedel06b} for a summary, 
\citealt{white01}). TW Hya is plotted twice, as two different accretion 
rates have been reported (see \citealt{kastner02}): from H$\alpha$
measurements, \citet{muzerolle00} derived $\dot{M} \approx
5 \times 10^{-10}$~M$_{\odot}$ yr$^{-1}$, while according to the excess 
in the C\,{\sc iv} $\lambda$1549 line and the empirical relation between this 
excess and the accretion rate described by \citet{johnskrull00}, 
$\dot{M}$ could reach values of $(3-6) \times 10^{-8}$~M$_{\odot}$ yr$^{-1}$.
We found no information on an upper limit of $\dot{M}$ for TWA 5 in
the published literature, and again recall that this stellar
system contains four components.

Despite the low statistics, a trend is clearly visible: stars with
low (or vanishing) accretion rates have $T_{\rm oxy}$ higher than 
the accreting stars. Observationally, these stars thus reveal a low 
O\,{\sc vii}/O\,{\sc viii} flux ratio.

\subsection{X-rays from accretion?}

\citet{kastner02} and \citet{stelzer04} have interpreted the low $f/i$ flux ratio
(consistent with high electron density),
and other properties of the X-ray spectrum of TW Hya, such as the low X-ray temperature and the
low Fe/Ne abundance ratio, in terms of accretion.  An analogous
scenario has been suggested by \citet{schmitt05}, \citet{robrade06}, and \citet{guenther06} to explain
the low $f/i$ ratio measured in three other stars, BP Tau, CR Cha, and V4046 Sgr. In the latter
two cases, however, the accretion shock would be responsible only for the softest component
of the spectrum, while magnetic activity or star-disk interaction is required to explain
the hotter components.

In our stellar sample, an O\,{\sc vii} triplet can be measured for only two stars: the CTTS
BP Tau and the Herbig star AB Aur.
For the two accreting stars SU Aur and DH Tau, the O\,{\sc vii} triplet is not present,
presumably because it is more absorbed ($N_{\rm H} = [3.1-3.2] \times 10^{21}$ cm$^{-2}$
and $N_{\rm H} = 2.0 \times 10^{21}$ cm$^{-2}$, respectively) and perhaps
also because the X-ray emission was dominated by flaring plasma.

The low $f/i$ flux ratio measured in BP Tau is consistent with
high densities ($n_{\rm e}= 3.4 \times 10^{11}~$cm$^{-3}$).
On the other hand, the $f/i$ ratio measured in AB Aur strictly excludes
high densities ($n_{\rm e} \la 10^{10}~$cm$^{-3}$).

We now test the accretion hypothesis for BP Tau and AB Aur.
The accretion rate of BP Tau is $\dot{M}_{\rm acc} \approx (1.32-2.88)
\times 10^{-8} M_{\odot}$~yr$^{-1}$ \citep{white01,muzerolle98}. For 
AB Aur we approximate the accretion rate as $\approx 10^{-8\pm 1} M_{\odot}$~yr$^{-1}$,
according to \citet{telleschi06a} and references therein.
The accretion luminosity, neglecting viscous dissipation, is given by
$L_{\rm acc}=G M \dot{M}/(2R)$, which can be written as $L_{\rm acc,30}
\approx 600 \tilde{M} \dot{M}_{-8} / \tilde{R}$, where $L_{\rm acc,30}=L_{\rm acc}/
(10^{30}$ erg s$^{-1}$), $\tilde{M}=M/M_{\odot}$, $\dot{M}_{\rm acc,-8}=\dot{M}_{\rm acc}/10^{-8}
M_{\odot}$~yr$^{-1}$  and $\tilde{R}=R/R_{\odot}$. 
Using the stellar parameters of Table~\ref{prop} we obtain, $L_{\rm acc,BP} \gtrsim 3 
\times 10^{32}~{\rm erg~s}^{-1}$ and $L_{\rm acc,AB} \approx 7 
\times 10^{32}~{\rm erg~s}^{-1}$, i.e. enough to account for the observed luminosities.

We further estimate the temperature expected for the accretion shock. The temperature
in case of strong shocks is given by $T = 3 v^{2} \mu m_{p}/16k$, where the velocity
is approximately the free fall velocity $v_{\rm ff}=(2GM/R)^{1/2}$, $m_{p}$ is the
proton mass, $k$ is the Boltzmann constant, and the mean molecular weight $\mu 
\approx 0.62$ for a fully ionized gas.
We thus find $T \approx 5.4 \times 10^{6} \tilde{M}/\tilde{R}$ [K], and with the
parameters from Table~\ref{prop}, $T_{\rm BP} \approx 2$ MK and $T_{\rm AB} \approx 6$ MK.
For BP Tau, this temperature is consistent with the temperature of the soft
component in the 3-$T$ fit and with the temperature derived from
Fig.~\ref{o_ratio}; for AB Aur, $T_{\rm AB}$ is consistent
with $T_{\rm av}$.

We can further estimate the shock density, using the strong-shock
condition $n_2$=4$n_1$, where $n_1$ and $n_2$ are the pre-shock 
and post-shock densities.
The density $n_1$ can be estimated from the accretion mass rate and the 
accreting area on the stellar surface: $\dot{M} \approx 4 \pi R^{2}fv_{\rm ff}n_em_p$,
where $f$ is the surface filling factor of the accretion flow. We
thus find
\begin{equation}
n_2 \approx {4 \times 10^{11}\over \tilde{R}^{3/2}\tilde{M}^{1/2}} {\dot{M}_{-8} \over f}~{\rm [cm^{-3}]}.
\label{eq.density}
\end{equation}
According to \citet{calvet98}, typical values for  $f$ are $f= 0.1-10$\%. The
density should then be $n_2 = 10^{12}-10^{14}$~cm$^{-3}$ for both stars, given the
adopted $\dot{M}$. 

These densities are therefore
compatible with the O\,{\sc vii} triplet fluxes measured in BP Tau, but not with
those in AB Aur. In order to obtain the electron density that
we measure in AB Aur from Eq.~\ref{eq.density}, $\dot{M}_{\rm AB}$ should be lowered to 
about $10^{-10} M_{\odot}$~yr$^{-1}$; also the accreting area should be at least
10\% and probably more, essentially
the whole stellar surface. The first possibility is not supported from the (tentative)
measurements of $\dot{M}$ (\citealt{telleschi06a} and references therein), while
a filling factor approaching unity is unreasonable, given that the star accretes from a disk
and a wind is present \citep{praderie86}. We also note that a low
$f/i$ ratio has also been measured for the CTTS T Tau \citep{guedel06a, guedel06c}.

A problem with the accretion scenario is that the shock is formed close
to the stellar photosphere \citep{calvet98} and that the X-rays could
therefore be absorbed. \citet{drake05a} studied the problem for TW Hya.
The depth of the shock can be estimated from the measured electron density.
For TW Hya the density of $n_{\rm e} \approx 10^{13}$~cm$^{-3}$, derived from 
Ne\,{\sc ix} and O\,{\sc vii} triplets, corresponds to a
larger $N_{\rm H}$ than observed. \citet{drake05a} suggested that $n_{\rm e}$ is 
only $\approx 10^{12}$~cm$^{-3}$ in TW Hya while photoexcitation from the ambient
UV radiation field could be responsible for the observed $f/i$ ratio.
For BP Tau, we measure an electron density smaller than in TW Hya, and we
expect therefore that the shock is higher in the photosphere or above the 
photosphere, and the problem of photospheric absorption is at least
in part alleviated, while the influence of the UV radiation field from
the shock on the $f/i$ ratio still remains unknown.

Accretion can therefore explain the soft excess that we measure in
BP Tau, but cannot explain the soft X-ray emission in AB Aur.

\section{Conclusions}\label{conclusions}

We have presented high resolution X-ray spectra of nine young stellar
objects. Five of them are accreting stars (four CTTS and one Herbig
star) and four are WTTS. From previous work on high-resolution X-ray 
spectroscopy of T Tauri stars \citep{kastner02, stelzer04,
argiroffi05, schmitt05, robrade06,guenther06}, three X-ray 
properties have been proposed to characterize accreting pre-main-sequence 
stars:
i) they show strong soft emission (TW Hya); 
ii) they display a high electron density (TW Hya, BP Tau,  V4046 Sgr); 
iii) the Ne or N abundances are relatively high (TW Hya, BP Tau, V4046 Sgr),
when compared with Fe. However, the sample of stars that had been studied
so far was too small to prove that these three properties are common to all
CTTS. Our sample adds a significant number of spectra to 
test these conjectures.

For two accreting stars (BP Tau and AB Aur), we have been able to measure
the O\,{\sc vii} triplet and to derive the source electron density. 
While we measured a high density for BP Tau (3.4$\times 10^{11}$~cm$^{-3}$,
confirming previous reports, \citealt{schmitt05}), 
the density for the Herbig star AB Aur is low, with  $n_{\rm e} <10^{10}$~cm$^{-3}$.
In the high-resolution X-ray spectrum of the CTTS T Tau (XEST-01-045), the O\,{\sc vii} triplet is
also consistent with a low density, as reported by \citet{guedel06a}. The low 
signal-to-noise ratio does not allow to constrain the electron density for the CTTS CR 
Cha \citep{robrade06} and DN Tau sufficiently well. For the other CTTS, the 
O\,{\sc vii} triplet was not measurable because of strong absorption or, probably,
due to the dominance of hot material during strong flaring (in SU Aur and DH Tau).

Regarding WTTS, none of the WTTS in our sample displays an O\,{\sc vii} triplet,
testifying to the low flux in these lines despite their low absorption, and therefore 
pointing to a deficiency of cool material if compared with CTTS. The  O\,{\sc vii} 
triplet was previously  measured for two other WTTS, TWA~5 \citep{argiroffi05} and 
HD~98800 \citep{kastner04}, and in both cases relatively low densities,
reminiscent of the coronae of main-sequence stars, were observed. 

In conclusion, apart from several T Tauri stars for which the O\,{\sc vii} remains
undetected, there are so far two clear reports each for high (TW Hya, BP Tau) and 
for low densities (AB Aur, T  Tau) in the cooler plasma component.  Although 
high densities are not a condition to qualify the soft X-ray emission for the
accretion scenario, we suggested that the accretion rates and the filling factors 
would make this scenario unlikely for AB Aur. A specific discussion of the T Tau
observation will be given by \citet{guedel06c}. The triplet
line flux ratios do therefore not seem to be reliable indicators for 
accretion-induced X-rays on T Tau stars.

As for the issue of overabundances of specific elements such as Ne or N, we 
have indeed found the Ne abundance to be high compared
to Fe (4-6 times higher than the solar ratio), but this is the case for {\it all} stars
except the G type stars (SU Aur, HP Tau/G2, and HD 283572) and the Herbig star AB Aur. 
Stars with a high Ne abundance in their X-ray source thus comprise both WTTS and 
CTTS. A high Ne abundance has also been reported for three members of the TW association,
namely TW Hya \citep{kastner02}, HD~98800 \citep{kastner04}, and TWA~5 \citep{argiroffi05},
of which the latter two seem to be non-accreting T Tau stars. 
Therefore, we suggest that the high Ne abundance is not a characteristic property
of CTTS, but is common to most young low-mass stars. 
Studying the abundances of our stellar sample and a sample of other PMS
and active MS stars, we have found that the G-type stars on average show abundance 
ratios of $\langle$Fe/Ne$\rangle_{\rm G} = 1.02$ and $\langle$Fe/O$\rangle_{\rm G}=2.03$, 
while we have found significantly lower ratios for K-type stars (on average 
$\langle$Fe/Ne$\rangle_{\rm K} =0.22$ and $\langle$Fe/O$\rangle_{\rm K} =0.58$).
It thus seems that the abundance ratios are a function of the spectral type and are 
similar for PMS stars and more evolved active stars, while we find no trend with
respect to accretion.
The abundance of N
is difficult to measure because the relevant lines are usually rather faint or
suppressed by photoelectric absorption.  \citet{telleschi06a} do not find a 
significant anomaly in AB Aur for which the N\,{\sc vii} line could be fitted.

Finally, the outstanding property of the TW Hya spectrum reported previously
\citep{kastner02} is the dominance of soft emission relating to cool plasma.
Other T Tau stars, however, regularly show hard spectra from which a dominant
hot plasma component is inferred (e.g., \citealt{preibisch05, guedel06b}).
Here, we have again made specific use of the high-resolution available from the RGS
instrument at low photon energies. We have  studied the flux ratio of 
O\,{\sc vii}/O\,{\sc viii} that is inaccessible to CCD instruments. The O\,{\sc vii}
line is formed in a relatively narrow temperature range ($\approx 1-4$~MK) with
the emissivity peaking at $\approx 2$~MK, whereas
the   O\,{\sc viii} line forms over a wider range centered at somewhat higher temperatures
($\approx$ 4~MK). The  O\,{\sc vii}/O\,{\sc viii} flux ratio is thus a good temperature indicator 
for the plasma at the cool end of the emission measure distribution. The ratio
may be modified by photoelectric absorption, but we have corrected for this effect
and determined the single temperature that is equivalent to the unabsorbed flux ratio
(see Fig.~\ref{o_ratio}).  In our sample, it is evident
that the accreting stars show an {\it excess} in the softest emission 
(expressed by lower $T_{\rm oxy}$ in Fig.~\ref{o_ratio}) that is
not present in the non-accreting stars. While this {\it soft excess} may
be of little relevance for the overall emission measure distribution dominated
by plasma up to 30~MK, it does systematically alter the lines formed at the lowest temperatures.
The excess of cool emission measure could be due to additional volume containing
cool plasma, or due to increased densities of low-temperature plasma. We cannot 
conclusively distinguish between these alternatives, but note that accreting 
stars with high and low densities as inferred from O\,{\sc vii} line ratios have now been 
reported. We conclude that while hot plasma may dominate the X-ray sources of most
T Tau stars, a {\it soft excess} is characteristic of the accreting stars.

What, then, is the mechanism that induces a soft excess in the accreting 
subsample of our targets? The accretion shock scenario \citep{kastner02} remains 
a possibility for TW Hya, BP Tau, and V4046 Sgr although the requirements for AB Aur and
T Tau are rather demanding, as lower accretion rates than hitherto estimated
and/or very large accretion areas on the star are required, in contradiction to
the standard magnetic-funnel accretion scenario \citep{calvet98}.

\citet{audard05} have observed a strongly accreting T Tau star (a so-called
EXor object) during an outburst attributed to a strong accretion event. They
noted a significant softening of the spectrum during and after outburst, indicating
the predominance of cooler plasma, although the X-ray luminosity did not change
significantly. They  speculated that the accreting material is disrupting the largest
magnetic features during outburst, which would also be the hottest if magnetic 
loops have the  same pressure. This would favor emission from smaller, cooler
magnetically confined regions in those areas where accretion is active.

Alternatively, the accreting material may not disrupt the magnetic structures but
may rather fill them with additional cool material that is not driven into the
coronal regions from the chromospheric layers by the mechanism of coronal heating.
The plasma in the accreting loops is thus cooler from the outset. Also, the 
increased electron density increases the cooling efficiency of heated loops,
because the cooling losses scale as $n_{\rm e}^2$. Magnetic loops loaded
with accreting material are therefore cooler \citep{preibisch05} if not otherwise
heated preferentially.

\begin{acknowledgements}

We thank Beate Stelzer for helpful suggestions and information on TW Hya, and Laurence DeWarf, Edward 
Fitzpatrick, and Claude Catala 
for important information on fundamental properties of AB Aur.
We also thank Joel Kastner for information on his TW Hya and HD 98800 spectral analysis.
An anonymous referee provided helpful comments that have significantly improved
the paper.
This research is based on observations obtained with {\it XMM-Newton}, an ESA science
mission with instruments and contributions directly funded by ESA Member States and
the USA (NASA).
We would like to thank the International Space Science Institute (ISSI) in Bern,
Switzerland, for logistic and financial support during several workshops on the TMC campaign.
X-ray astronomy research at PSI has been supported by the Swiss National Science
Foundation (grant 20-66875.01 and 20-109255/1). M.A. acknowledges support from NASA grant NNG05GF92G.
In addition, he acknowledges support from a Swiss National Science Foundation Professorship (PP002--110504).

\end{acknowledgements}
%
%

\end{document}